\documentclass[aps,prd,twocolumn,twoside,superscriptaddress,floatfix, nofootinbib]{revtex4-1}
\pdfoutput=1
\usepackage{ifthen,amsmath,amssymb, bm}
\usepackage{graphicx}
\usepackage[dvipsnames]{xcolor}
\usepackage{hyperref}
\usepackage{float}
\usepackage{aas_macros}
\usepackage{caption}
\usepackage{subcaption}
\usepackage[export]{adjustbox}
\usepackage{bookmark}

\newcommand{\beq} {\begin{equation}}
\newcommand{\eeq} {\end{equation}}
\newcommand{\bal} {\begin{aligned}}
\newcommand{\eal} {\end{aligned}}

\begin{document}

\title{Velocity reconstruction in the era of DESI and Rubin (part I):\\ 
Exploring spectroscopic, photometric \& hybrid samples
}

\author{Bernardita Ried Guachalla}
\email{bried@stanford.edu}
\affiliation{Department of Physics, Stanford University, Stanford, CA, USA 94305-4085}
\affiliation{Kavli Institute for Particle Astrophysics and Cosmology, 382 Via Pueblo Mall Stanford, CA 94305-4060, USA}
\affiliation{SLAC National Accelerator Laboratory 2575 Sand Hill Road Menlo Park, California 94025, USA}
\author{Emmanuel Schaan}
\affiliation{Kavli Institute for Particle Astrophysics and Cosmology,
382 Via Pueblo Mall Stanford, CA 94305-4060, USA}
\affiliation{SLAC National Accelerator Laboratory 2575 Sand Hill Road Menlo Park, California 94025, USA}
\author{Boryana Hadzhiyska}
\affiliation{Miller Institute for Basic Research in Science, University of California, Berkeley, CA, 94720, USA}
\affiliation{Physics Division, Lawrence Berkeley National Laboratory, Berkeley, CA 94720, USA}
\affiliation{Berkeley Center for Cosmological Physics, Department of Physics, University of California, Berkeley, CA 94720, USA}
\author{Simone Ferraro}
\affiliation{Physics Division, Lawrence Berkeley National Laboratory, Berkeley, CA 94720, USA}
\affiliation{Berkeley Center for Cosmological Physics, Department of Physics, University of California, Berkeley, CA 94720, USA}

%%%%%%%%%%%%%%%%%%%%%%%%%%%%%%%%%%%%%%%%%%%%%%%%%%%%%%%
\begin{abstract}

%what/context/motivation
Peculiar velocities of galaxies and halos can be reconstructed from their spatial distribution alone.
This technique is analogous to the baryon acoustic oscillations (BAO) reconstruction, using the continuity equation to connect density and velocity fields.
The resulting reconstructed velocities can be used to measure imprints of galaxy velocities on the cosmic microwave background (CMB) like the kinematic Sunyaev-Zel'dovich (kSZ) effect or the moving lens effect.
As the precision of these measurements increases, characterizing the performance of the velocity reconstruction becomes crucial to allow unbiased and statistically optimal inference.
% this paper/goal
In this paper, we quantify the relevant performance metrics: the variance of the reconstructed velocities and their correlation coefficient with the true velocities.
% halo VS galaxy velocities
We show that the relevant velocities to reconstruct for kSZ and moving lens are actually the halo -- rather than galaxy -- velocities.
% various parameters we change
We quantify the impact of redshift-space distortions, photometric redshift errors, satellite galaxy fraction, incorrect cosmological parameter assumptions and smoothing scale on the reconstruction performance.
We also investigate hybrid reconstruction methods, where velocities inferred from spectroscopic samples are evaluated at the positions of denser photometric samples.
We find that using exclusively the photometric sample is better than performing a hybrid analysis.
% method
The 2 Gpc$/h$ length simulations from \textsc{AbacusSummit} with realistic galaxy samples for DESI and Rubin LSST allow us to perform this analysis in a controlled setting.
% companion paper
In the companion paper \cite{Hadzhiyska2023}, we further include the effects of evolution along the light cone and give realistic performance estimates for DESI luminous red galaxies (LRGs), emission line galaxies (ELGs), and Rubin LSST-like samples.

\end{abstract}
\maketitle

%%%%%%%%%%%%%%%%%%%%%%%%%%%%%%%%%%%%%%%%%%%%%%%%%%%%%%%
\section{Introduction}

Astronomical observations are not simply static snapshots of the Universe.
While photometric and spectroscopic observations give us the instantaneous positions of objects on the sky (2D) and sometimes their distances (3D),
their velocities contain invaluable clues to the gravitational potentials that cause them.
On the shortest scales, solar neighborhood objects' orbits \cite{Poincare1899, Arnold1976} tell us about the masses of their attractors.
On galactic scales, the proper motion of nearby and distant stars using parallaxes \cite{Perryman1997, Bailer_Jones_2015} or intrinsic properties (e.g. Cepheids stars \cite{Leavitt1912}) give us the gravitational structure of the Milky Way. 
At larger distances, the ``Hubble flow'' of the Universe's expansion appears to us as a cosmological redshift in galaxy spectra \cite{Lemaitre1927, Hubble1929, Sandage1958, Riess_1995}.

Galaxy motions beyond the Hubble flow, the so-called peculiar velocities \cite{Kaiser1987, Strauss1995, Howlett_2022}, are key probes of the instantaneous growth rate of matter fluctuations, i.e. the dynamics of the large-scale structure.
Comparing them with static probes like galaxy clustering or gravitational lensing enables key tests of general relativity \cite{Percival_2009}.
Peculiar velocities can also be a powerful probe of the matter density field on the largest scales, outperforming galaxy clustering in some regimes.
Indeed, measuring mass by counting galaxies is limited by the finite number of galaxy available, which translates into galaxy shot noise.
On the other hand, knowing the position and velocity of a single galaxy perfectly can tell you exactly the mass of the object it is circling around
\cite{smith2018ksz, Munchmeyer_2019, Peebles_2022}\footnote{As expressed in \cite{Peebles_2022}, because ``we see that in the gravity physics of the standard cosmology a mass concentration that is in principle too far away to be observed can produce a flow that in principle can be observed. \cite{Grishchuk1978} may have been the first to recognize this."}.
Thus peculiar velocities can significantly improve measurements of local primordial non-Gaussianity ($f_{\rm NL}^{\rm local}$), via sample variance cancellation on the scale-dependent bias \cite{smith2018ksz, Munchmeyer_2019}.

Peculiar velocities can be observed as Doppler shifts in galaxy spectra.
They can be distinguished from the cosmological redshifts for nearby galaxies whose distances are known from Tully-Fisher \cite{Tully1977}, fundamental plane \cite{1987Djorgovski} relations or supernovae type Ia \cite{Branch1993}. 
%\SF{Perhaps let's mention Supernovae 1a since it's becoming a large field}.
In the more distant Universe, peculiar velocities appear to us as redshift-space distortions (RSD) in the 3D galaxy positions inferred from their redshifts, and imprint key signatures on their clustering.

However, galaxy and dark matter peculiar velocities also leave observable imprints on the most distant source of our observable universe: the cosmic microwave background (CMB).
As moving electrons Thomson scatter off CMB photons, they imprint their line-of-sight (LOS) velocity on the CMB in the form of the kinematic Sunyaev-Zel'dovich (kSZ) effect  \cite{Sunyaev1980}.
This effect also gives us fundamental information about the smaller scales of the LSS, by localizing the ``missing'' baryons \cite{Fukugita_2004, Bullock2017, Battaglia_2017}.
Indeed, the baryon abundance inferred from early-Universe observations (including the primary CMB) is higher than estimated from the galaxy starlight \cite{Walker1991, White1993, Planck2020}.
The ``missing'' baryons, in the form of the warm-hot intergalactic medium (WHIM) \cite{Cen_2006} around galaxies, can be revealed in multiple ways (e.g., X rays \cite{1998Bryan, 2009Sun}, fast radio bursts \cite{2020Macquart} and quasar absorption lines \cite{2018Nicastro}).
Among these, kSZ stands out by measuring baryon density directly (without needing to model gas temperature) in the outskirts (beyond the virial radius) of low-mass halos (group-sized) at high redshift.
Localizing the baryons in these environments can lift one of the most limiting systematic in galaxy lensing studies from the Rubin Observatory Legacy Survey of Space and Time (LSST) \cite{lsstsciencecollaboration2009lsst}, the Euclid space telescope \cite{Amendola_2018} and the Nancy Grace Roman Space Telescope \cite{Spergel_2015}.

Furthermore, dark and ordinary matter in moving halos cause time-varying gravitational potentials.
CMB photons crossing these changing potentials will see a net energy gain or loss, analogous to the integrated Sachs-Wolfe (ISW) effect.
This so-called moving lens effect \cite{Birkinshaw1983, Hotinli2019, Hotinli2021} imprints the halo transverse peculiar velocities on the CMB.
This effect may be the only viable probe of the transverse motion of distant galaxies, and will be detectable with the upcoming CMB ground-based experiments like the Simons Observatory \cite{Ade2019} and CMB-S4 \cite{Abazajian2016} cross-correlated with galaxy surveys such as the Dark Energy Spectroscopic Instrument (DESI) \cite{Aghamousa2016} and LSST \cite{lsstsciencecollaboration2009lsst}.

Thus CMB maps contain information about all three components of galaxy and halo peculiar velocities.
To extract this information, external handles on the 3D peculiar velocities help separate the kSZ and moving lens signals from the primary CMB and other sources of noise in CMB maps.
In this paper, we use the velocity reconstruction from the galaxy number density field, highly analogous to the baryon acoustic oscillations (BAO) reconstruction.
This method adopts a first-order (Zel'dovich) approximation \cite{Zeldovich1970} and solves the continuity equation of matter in redshift space \citep{Eisenstein_2007, Padmanabhan_2012} to infer velocities from the density field.
However, understanding the accuracy and precision of this method is crucial to provide optimal and unbiased inference of the kSZ and moving lens effects.

In this paper, we therefore assess the performance of the velocity reconstruction by using realistic simulations of galaxy samples resembling the DESI luminous red galaxies as well as LSST galaxies.
We focus on a simple periodic cubic box, in order to deconstruct the impacts of redshift-space distortions (RSD), the number density of galaxies in the sample, their satellite fraction, and their potential photometric redshift errors.
We also assess the performance of a hybrid spectro-photometric reconstruction, where the 3D velocity field is reconstructed from sparser spectroscopic galaxies, then evaluated at the positions of denser photometric galaxies.

In the companion paper \cite{Hadzhiyska2023}, we include further realism by considering the effect of evolution along the light cone in the analysis and realistic footprint masks, in order to provide the most realistic performance estimate for the velocity reconstruction, when applied to DESI-like and LSST-like galaxies.
These performance metrics will be a direct input to upcoming kSZ stacking measurements from DESI and the Atacama Cosmology Telescope (ACT) \cite{Fowler2007, Swetz2011, Thornton2016}.

This paper is organized as follows: 
In Section \ref{sec:methodology}, we present our methodology, including the velocity reconstruction mechanism, the simulated mocks, and the argumentation on reconstructing halo velocities rather than galaxy velocities.
In the following Section \ref{sec:results}, we explore the impact of changing different variables, such as the effect of photo-$z$ uncertainties, the number density, the satellite fraction, a different cosmology from the fiducial value and incorporating photo-$z$ galaxies in a spectroscopic sample.
We present our conclusions in Section \ref{sec:conclusions}.

%%%%%%%%%%%%%%%%%%%%%%%%%%%%%%%%%%%%%%%%%%%%%%%%%%%%%%%
%%%%%%%%%%%%%%%%%%%%%%%%%%%%%%%%%%%%%%%%%%%%%%%%%%%%%%%

\section{Simulating velocity reconstruction on DESI LRGs}
\label{sec:methodology}

In this section, we present the simulations and galaxy mocks used in this paper and the companion paper \cite{Hadzhiyska2023}, along with the 3D velocity reconstruction algorithm.
We present the key relevant performance metrics, and why the goal for kSZ and moving lens should be to reconstruct halo velocities, rather than galaxy velocities.

%%%%%%%%%%%%%%%%%%%%%%%%%%%%%%%%%%%%%%%%%%%%%%%%%%%%%%%

\subsection{\textsc{AbacusSummit} box simulation}
\label{sec:simulation}

We use one of the simulated boxes from \textsc{AbacusSummit} \cite{Maksimova_2021}.
These are public high-accuracy N-body simulations produced with the \textsc{Abacus} code \cite{Garrison2019, Garrison2021} that consists of 150 simulation boxes, spanning 97 cosmological models and different resolutions. 
The \textsc{AbacusSummit} suit of simulations meet all the Cosmological Simulation Requirements of the DESI survey \cite{DESIReport}, including the volume needed to properly sample the large-scale peculiar velocities, and the mass resolution needed for DESI-like galaxies. 

We focus on the simulated box \texttt{AbacusSummit\_base\_c000\_ph000} \cite{Maksimova_2021}, which follows the latest Planck $\Lambda$CDM cosmology\footnote{$\Omega_b h^2 = 0.02237$, $\Omega_c h^2 = 0.12$, $h = 0.6736$, $10^9 A_s = 2.0830$, $n_s = 0.9649$, $w_0 = -1$, $w_a = 0$} \cite{Planck2020}.
This box has a length of 2 Gpc/$h$, with a total of 6912$^3$ particles, each with a mass of $M_{\rm part} = 2.1 \times 10^9 M_\odot / h$.

Specifically, we use the $z=0.5$ snapshot, with $z$ being the redshift, as one of the well-explored redshift epochs of the Baryon Oscillation Spectroscopic Survey (BOSS) \cite{Eisenstein_2011, Dawson_2012}, DESI and LSST surveys.
Table \ref{tab:fiducial_parameters} summarizes the fiducial parameters assumed for cosmology, galaxies and the reconstruction method.
\begin{table}
	\centering
	\begin{tabular}{|c|c|}
		\hline
            Parameter & Value \\
            \hline
            $h$ & 0.6736 \\
            $\Omega_m$ & 0.31519 \\
            $b$ & 2.2 \\
            \hline
            $\sigma_z^{\rm DESI}$ & 0.00  \\
            $\sigma_z^\text{LSST goal}$ & $0.02 \cdot (1+z)$ \\
            $\sigma_z^\text{LSST requirement}$ & $0.05 \cdot (1+z)$ \\
            $\overline{n}$ & $2.14 \cdot 10^{-4}$ Mpc$^{-3}$ \\
            $f_{\rm sat}$ & 0.12  \\
            \hline
            $r_s$  & 14.8 Mpc \\
            \hline
	\end{tabular}
     \caption{
    Cosmological, galaxy and reconstruction fiducial values varied to explore the impact of the velocity reconstruction method.
    In the first row, we present the photo-$z$ uncertainties $\sigma_z$ from experiments like DESI and LSST (goal and requirement). 
    The mean number density $\overline{n}$ is defined to follow the DESI requirements \cite{2023Zhou} and the satellite fraction $f_{\rm sat}$ is derived from implementing the \textsc{AbacusHOD} \cite{Yuan2023}.
    The smoothing radius $r_s$ indicates the scale in the reconstruction algorithm that vanishes small scales.
    Finally, the cosmological parameters $h$, $\Omega_m$ and linear bias $b$ are fixed to \cite{Planck2020} and \cite{2010Tinker} results respectively. 
    }
\label{tab:fiducial_parameters}
\end{table}

%%%%%%%%%%%%%%%%%%%%%%%%%%%%%%%%%%%%%%%%%%%%%%%%%%%%%%%

\subsection{Mock DESI LRG sample}
\label{sec:galaxy_sample}

The \textsc{AbacusSummit} N-body simulations used in this work and in our companion paper were populated with galaxies through the extended halo occupation distribution (HOD) code, \textsc{AbacusHOD} \cite{Yuan_2021, Yuan2023}.
The code is suitable for creating mock catalogs of different tracers, including emission-line galaxies (ELGs) and Luminous Red Galaxies (LRGs), and more information on the modeling and fits to observations can be found in \cite{Hadzhiyska2022a, Yuan_2021}.
In this work, we focus on a cubic box populated exclusively with LRGs at a fixed redshift of $z=0.5$, for which the mean number density corresponds to $\overline{n}(z=0.5)=2.14 \cdot 10^{-4}$ Mpc$^{-3}$ \cite{2023Zhou} and the bias to $b(z=0.5) = 2.2$ \cite{2010Tinker}.
See Fig. 1 and Fig. 3 from \cite{Hadzhiyska2023} and \cite{Yuan_2023_Placeholder} for the general response to redshift of $\overline{n}(z)$ and $b(z)$ respectively. 

In this work, we use the standard HOD model presented in \cite{2007Zheng}, in which the main relevant property is the mass of the central halo.
This HOD model splits the galaxies into centrals and satellites, and also dictates their positions and velocities.
\begin{figure}[H]
    \centering
    \includegraphics[width=0.48\textwidth]{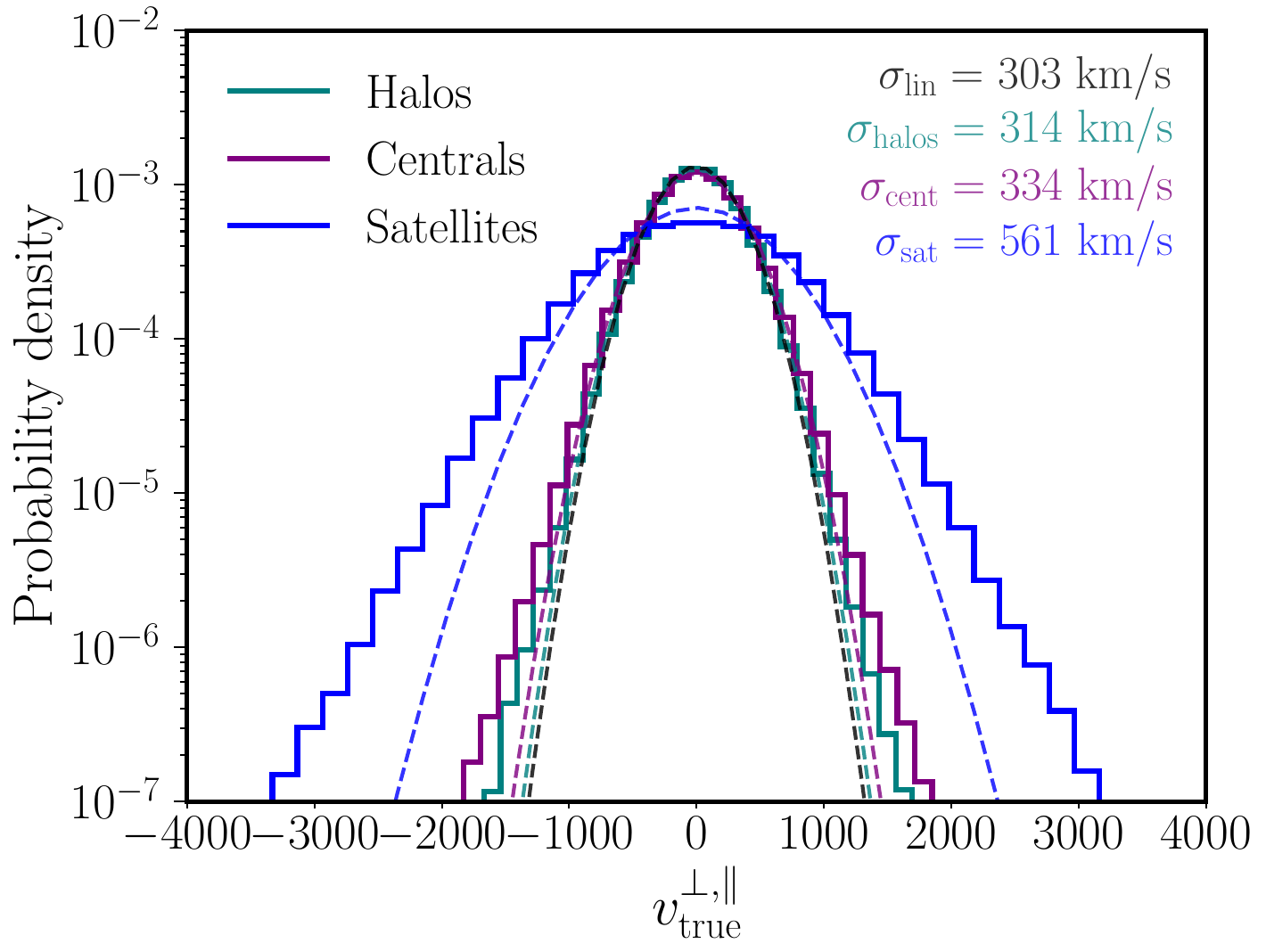}
    \caption{
    Histogram of the 1D velocities of the halos (teal) and galaxies (central (purple) and satellite (blue)) from the mock.
    To quantify deviations from a Gaussian due to non-linear evolution, normal distributions with the same standard deviations are shown in dashed.
    The standard deviation $\sigma_{\rm lin}$ corresponds to the expected cosmological standard deviation of the 1D velocity of matter Eq.~\eqref{eq:th_vel_std_dev}.
    The visible tails at large velocities are due to nonlinear evolution, including the pronounced virial motions of satellite galaxies.
    }
    \label{fig:vrec_vcent_vsat_1D_histogram}
\end{figure}

In particular, the velocity of the central galaxies is approximated as the velocity of the halo center \cite{Yuan2023} with a central velocity bias, while the satellites were randomly assigned to the positions and velocities of halo particles with uniform weights and also include a satellite velocity bias. 
The inclusion of velocity bias on central or satellite galaxies is motivated by observational results from spectroscopic surveys \cite{2015Guo} and simulations \cite{2017Ye}.
For central galaxies, there is a linear central velocity bias:
\begin{equation}
    v_{\rm cen} = v_{\rm halo} + \alpha_c \cdot \delta v(\sigma_{\rm halo})
    \label{eq:alpha_c}
\end{equation}
where $v_{\rm cen}$ is the central biased velocity, $v_{\rm halo}$ is the central subhalo velocity, $\alpha_c$ is the central velocity bias parameter and is restricted to be $\alpha_c \geq 0$ and $\delta v(\sigma_{\rm halo})$ is a Gaussian scatter.
For satellite galaxies, the velocity distribution is modulated as follows:
\begin{equation}
    v_{\rm sat} = v_{\rm halo} + \alpha_s \cdot (v_{\rm particle} - v_{\rm halo})
    \label{eq:alpha_s}
\end{equation}
where $v_{\rm sat}$ is the satellite biased velocity, $v_{\rm particle}$ denotes the corresponding dark matter particle velocity and $\alpha_s$ is the satellite velocity bias parameter \cite{Yuan_2021}.
It is worth noting that $\alpha_c$ = 0 and $\alpha_s$ = 1 mean no velocity bias for centrals and satellites respectively. 
We adopt the values $\alpha_c$ = 0.308 and $\alpha_s$ = 0.913 reported in Table 1 of \cite{hadzhiyska2023synthetic}, obtained fitting the \textsc{AbacusHOD} on the DESI SV3 (Survey Validation 3) LRGs data.

The second effect included in the HOD model corresponds to redshift-space distortions along the LOS, which in our case corresponds to the $z$ direction of the simulation box \cite{Yuan2023}. 
Based on those results, the \textsc{AbacusHOD} code fits the satellite fraction of LRGs to $f_{\rm sat} = 12\%$ \cite{yuan2023desi}.
For more details on how we obtained the HOD parameters for the DESI-like LRG samples see our companion paper \cite{Hadzhiyska2023}.

The resulting distributions of 1D peculiar velocities for halos and galaxies (centrals and satellites) are shown in Fig.~\ref{fig:vrec_vcent_vsat_1D_histogram}.
It is compared with the expectation for peculiar velocities of the matter field, estimated from the linear matter power spectrum as:
\beq
\sigma_\text{lin}^2
\equiv
\left( a H f \right)^2 \frac{2}{3}
\int 
\frac{dk}{(2\pi)^2}
\frac{P^\text{lin}(k)}{k^2}
.
\label{eq:th_vel_std_dev}
\eeq
where $P^\text{lin}(k)$ stand for the linear matter power spectrum calculated using CLASS \cite{Diego_Blas_2011}\footnote{
Integrating in the range of $k \in [10^{-5}, 3]$ Mpc$^{-1}$, we obtain a difference with respect to the non-linear case (using the halofit model \cite{Takahashi_2012}):
$\sigma_\text{lin} = 303$ km/s and $\sigma_\text{non-lin} = 342$ km/s respectively.
}.
In Sec. \ref{sec:perf_metrics} we comment on the effects of assuming this for the purposes of estimating the baryonic content of the universe through the kSZ effect.
For central galaxies, high-velocity tails appear due to the linear central velocity bias. 

For the purpose of velocity reconstruction, where exclusively large scale modes of the matter density field contribute, the only relevant difference between centrals and satellites is the amplitude of their random motions, leading to fingers of God. 
Indeed, the positional offsets of galaxies within a halo are a order of $\sim$Mpc, much smaller than the relevant scales for velocity reconstruction (10-100 Mpc).
Therefore, we are not going to vary the velocity bias parameters, as we will be varying the satellite fraction, which has a similar effect.

%%%%%%%%%%%%%%%%%%%%%%%%%%%%%%%%%%%%%%%%%%%%%%%%%%%%%%%

\subsection{Linear velocity reconstruction algorithm}
\label{sec:vel_rec}

The peculiar velocities of galaxies and their host halos can be estimated from the galaxy number density field.
Using the linear bias approximation \cite{Desjacques2018}, 
the galaxy number overdensity $\delta_g$ can be related to the underlying matter overdensity $\delta$ via
$\delta = b \delta_g$.
The continuity equation then connects the matter overdensity field to the peculiar velocity field $\mathbf{v}$:
for an overdensity of matter to be present at some position, matter must have been flowing towards that point.
To linear order in perturbations, and including the effect of redshift-space distortions (RSD), the continuity equation takes the form:
\begin{equation}
    \nabla \cdot \mathbf{v} + \frac{f}{b} \nabla \cdot [(\mathbf{v} \cdot \hat{\mathbf{n}}) \hat{\mathbf{n}}] = -a H f \frac{\delta_g}{b},
    \label{eq:continuity_eq}
\end{equation}
where $a$ is the scale factor, $H$ is the Hubble parameter, $\hat{\mathbf{n}}$ is the direction along the LOS, and $f$ is the logarithmic growth rate of structure
$f = d \ln D(a) / d \ln a$,
with $D(a)$ the linear growth factor.

The standard method of BAO reconstruction \cite{Eisenstein_2007, Padmanabhan_2012} solves the same equation in terms of the Lagrangian displacement, which can then be simply converted to velocity.
We therefore adapt the \texttt{MultiGrid} implementation of \cite{White2015} via the package \texttt{pyrecon}\footnote{\url{https://github.com/cosmodesi/pyrecon}}. 
Schematically, one can obtain the reconstructed velocity field via the following steps
\begin{enumerate}
    \item \textbf{Smoothed tracer number density field:} the galaxies are assigned to a 3D grid, and smoothed with a Gaussian filter $W(k, r_s) = \exp{[-\frac{1}{2} k^2 r_s^2]}$, where $k$ is the Fourier wavevector and $r_s$ the comoving smoothing length, in order to downweight modes dominated by shot noise.
    
    \item \textbf{Reconstructed displacement field:}
    In the plane parallel approximation, where the LOS direction is fixed throughout the box, an exact expression for the 3D Lagrangian displacement $\pmb{\psi}$ is then
    \begin{equation}
        \pmb{\psi}(\mathbf{k}) 
        = - i \frac{\mathbf{k}}{k^2} 
        \frac{\delta_g(\mathbf{k})}{(b + f \mu^2)} 
        W(k, r_s). 
    \end{equation}
    This operation, performed in a 3D cubic box, solves Eq.~\eqref{eq:continuity_eq}.
    The division by the sum of the linear bias $b$ and the linear redshift-space factor $f\mu^2$ converts the tracer number overdensity to the  matter overdensity, to linear order, and undoes the Kaiser effect (linear RSD).
    As usual, $\mu$ is the LOS angle cosine 
    $\mu \equiv \hat{\mathbf{n}} \cdot \mathbf{k} / {k}$.    
    
    \item \textbf{Reconstructed velocities:}
    We inverse-Fourier transform the displacement field $\pmb{\psi}(\mathbf{k})$ and convert it to the 3D velocity field via
    \begin{equation}
        \mathbf{v}_{\rm rec}(\mathbf{x}) = aHf \pmb{\psi}(\mathbf{x}),
    \end{equation}
    and evaluate it at the positions of each galaxy to obtain individual reconstructed velocities. 
\end{enumerate}
\begin{figure*}
    \centering
    \begin{subfigure}[b]{0.35\textwidth}
         \centering
         \includegraphics[width=0.99\textwidth]{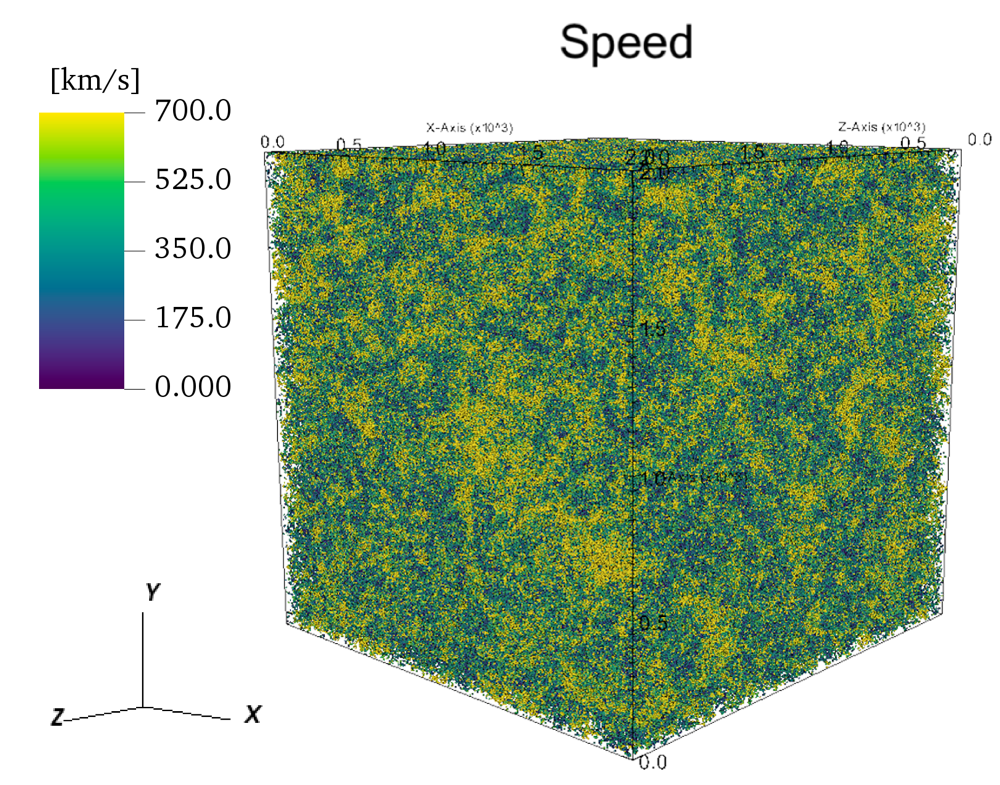}
    \end{subfigure}
    \hfill
    \hspace{-0.6cm}
    \begin{subfigure}[t]{0.35\textwidth}
         \centering
         \includegraphics[width=0.99\textwidth]{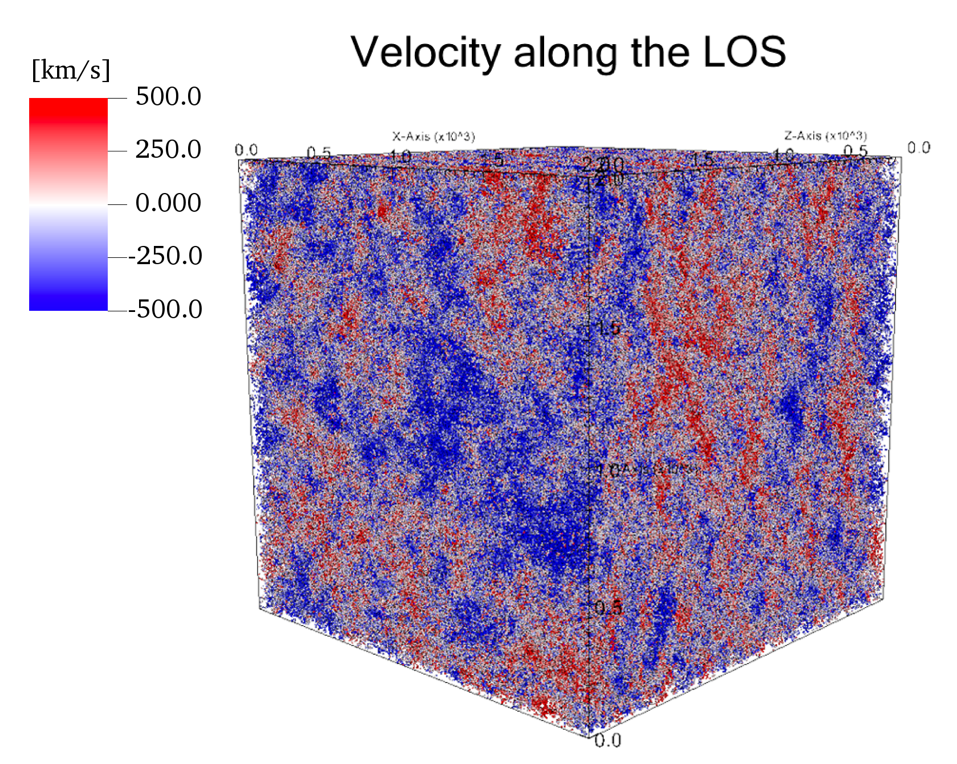}
         \caption{True}
    \end{subfigure}
    \hfill
    \hspace{-0.8cm}
    \begin{subfigure}[b]{0.34\textwidth}
         \centering
         \includegraphics[width=0.97\textwidth]{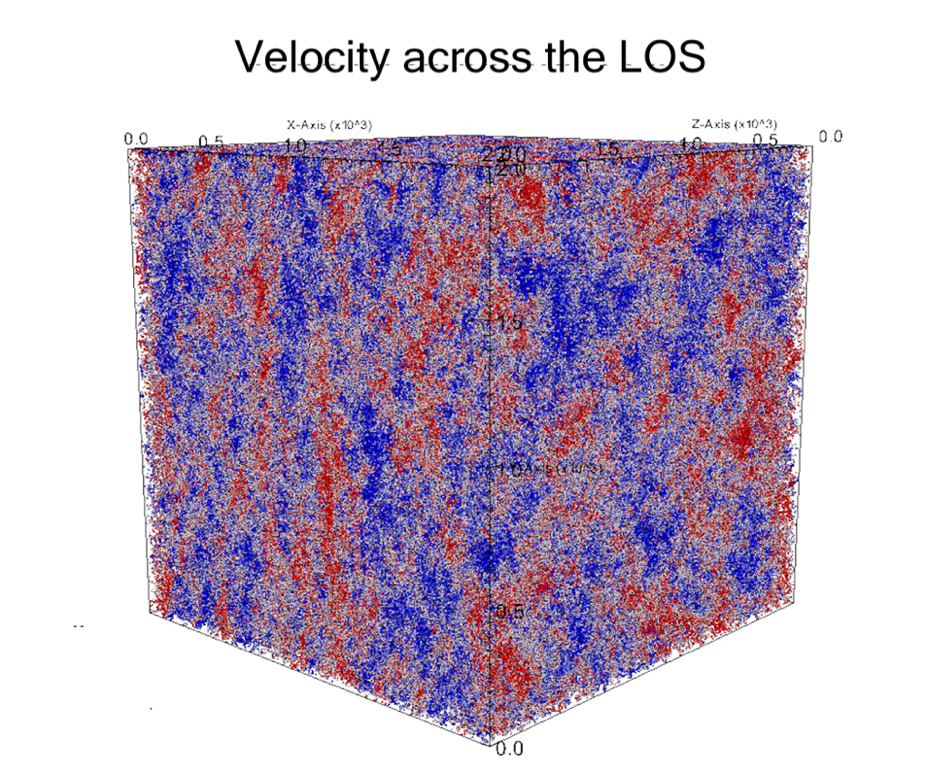}
    \end{subfigure}
    \hfill
    \vspace{-0.4 cm}

    \begin{subfigure}[b]{0.36\textwidth}
    \includegraphics[width=0.95\textwidth]{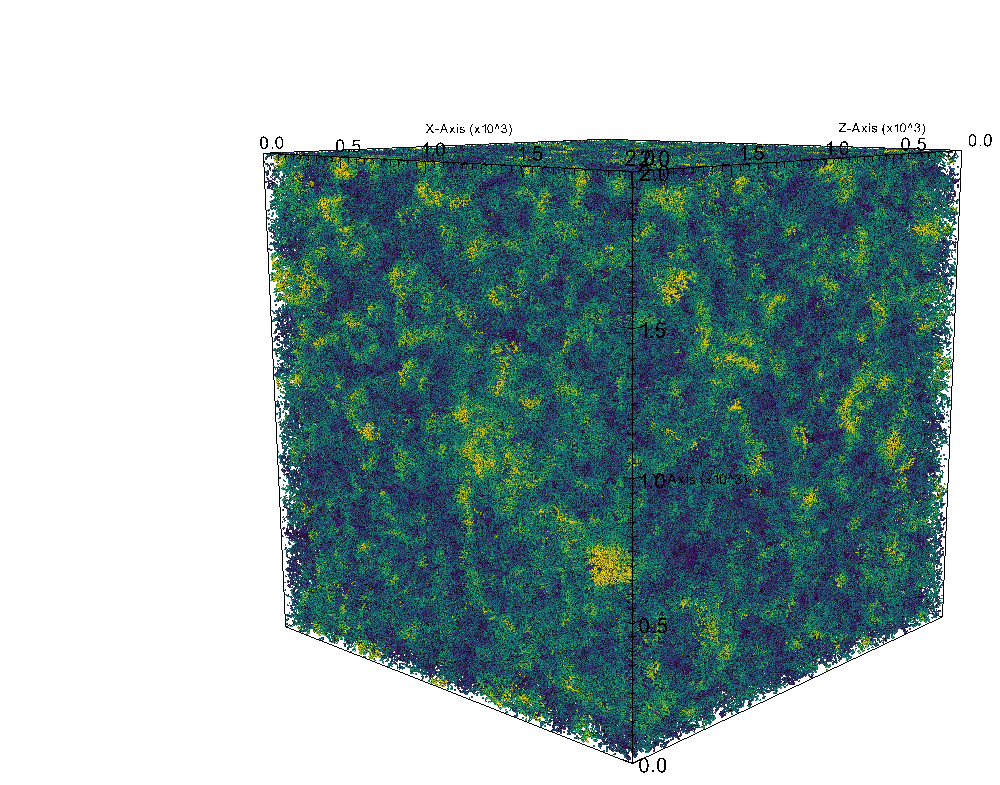}
    \end{subfigure}
    \hfill
    \hspace{-1.8cm}
    \begin{subfigure}[t]{0.32\textwidth}
         \centering
         \includegraphics[width=0.99\textwidth]{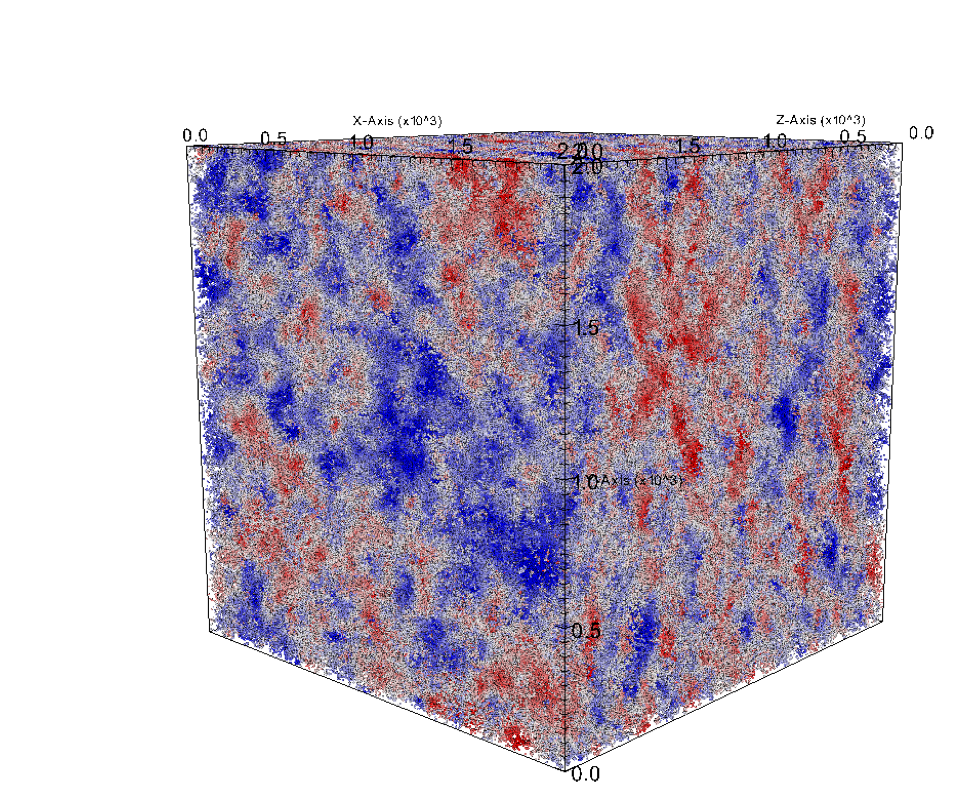}
         \caption{Reconstruction}
    \end{subfigure}
    \hfill
    \hspace{-1.9cm}
    \begin{subfigure}[b]{0.34\textwidth}
         \centering
         \includegraphics[width=0.99\textwidth]{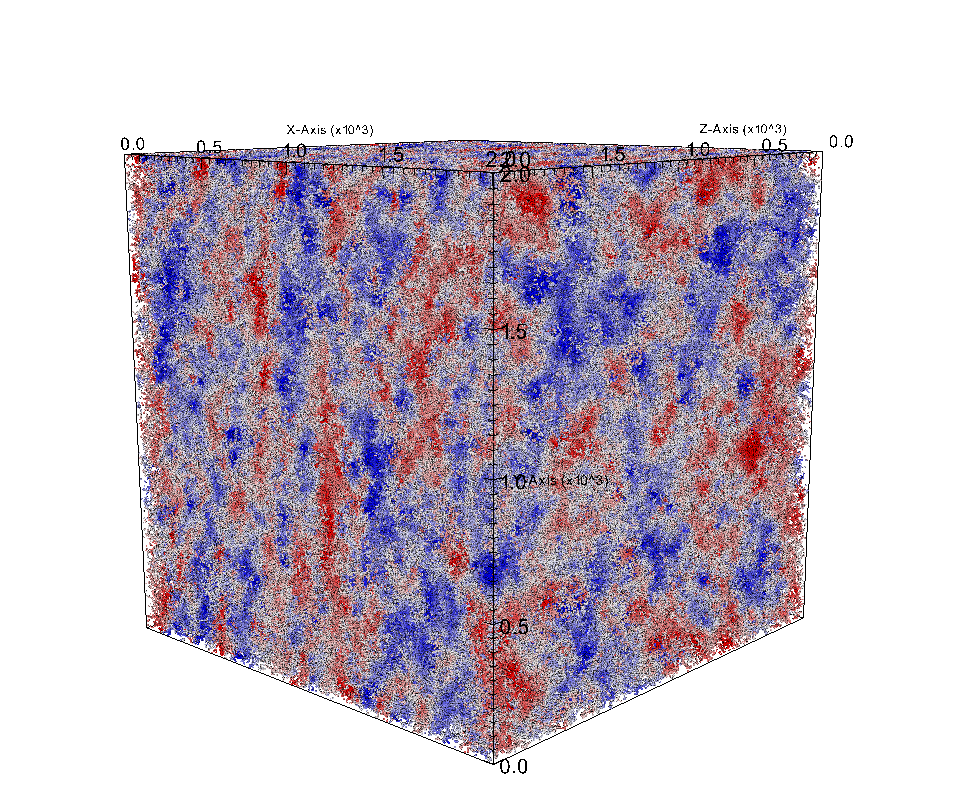}
    \end{subfigure}
    \caption{
    Visualization of velocity fields in the \textsc{AbacusSummit} simulation box at redshift $z$ = 0.5. 
    Top and bottom rows represent true and reconstructed velocities. 
    Line-of-sight (LOS) is defined as the $z$ axis (i.e. from lower left pointing towards the paper). 
    \textit{Left column}: Speed, i.e. modulus of the velocity.
    \textit{Center column}: LOS velocity component
    \textit{Right Column}: One of the components across the LOS ($x$ component; the $y$-component is statistically identical).
    The reconstructed velocities have the same large-scale structure as the true ones, though with a reduced modulus, as expected due to the linear approximation and smoothing scale.
    These visualizations also illustrate how the speed field is isotropic, whereas the 1D components have preferred directions: the elongated red and blue blobs indicate that the $j$-component of the velocity ($j=x, y, z$) varies fastest along the $j$-direction.
    This is predicted by linear theory, where the velocity field is the gradient of an isotropic field.
    }
    \label{fig:cubic_box_visualization}
\end{figure*}
In the left panel of Fig. \ref{fig:cubic_box_visualization} we show the true velocity field on the top and the reconstructed velocity field on the bottom for an \textsc{AbacusSummit} box populated with DESI LRGs at $z = 0.5$. 
The velocity structure on the largest scales is visibly well-reconstructed along and across the LOS.
However, the linear reconstruction underestimates the magnitudes of galaxy velocities, since it misses the non-linear Fourier modes.
Furthermore, it does not account for small scale RSD, like the Fingers of God (FoG) effect that is present in satellite galaxies. 

To assess the impact of RSD and potential photometric errors, we study separately the velocity reconstruction along and across the LOS as done in \cite{chan2023reconstructing}:
\begin{equation}
    \mathbf{v}_{\rm rec} = 
    \begin{pmatrix}
    v_{\rm rec}^{\perp}\\
    v_{\rm rec}^{\|}
    \end{pmatrix},
\end{equation}
and show $v^{\|}$ and $v^{\perp}$ respectively in the middle and right panels of Fig.~\ref{fig:cubic_box_visualization}, where the $z$ axis corresponds to the LOS direction.

%%%%%%%%%%%%%%%%%%%%%%%%%%%%%%%%%%%%%%%%%%%%%%%%%%%%%%%

\subsection{Relevant performance metrics for kSZ \& moving lens}
\label{sec:perf_metrics}

Our goal is to use the velocity reconstruction in measurements of the kSZ or moving lens effects.
In particular, the kSZ effect is a Doppler shift of the CMB photons due to scattering with the surrounding gas of galaxies and clusters moving along the LOS.
This effect impact the temperature of the CMB photons as follows:
\begin{equation}
    \frac{\delta T_{\rm kSZ}}{T_{\rm CMB}} 
    \sim
    - \int d\chi ~n_e \sigma_T ~(\mathbf{v}_{\rm pec} \cdot \hat{\mathbf{n}})
\end{equation}
where $\chi$ is the comoving distance, $n_e$ is the electron number density, $\sigma_T$ is the Thomson cross-section, $\mathbf{v}_{\rm pec}$ the peculiar velocity and $c$ the speed of light.
For kSZ, a typical estimator $\epsilon$ for stacking is of the form \cite{Schaan2021}: 
\beq
\bal
\epsilon 
&\propto \langle \delta T_\text{kSZ}\  v_\text{rec} \rangle \\
&\propto \langle n_e v_\text{gas}\  v_\text{rec} \rangle \\
&= n_e\ r\ \sigma_\text{gas} \sigma_\text{rec},
\eal
\label{eq:epsilon}
\eeq
Therefore, to convert a kSZ measurement into an electron number density profile, we need to know
\begin{itemize}
\item The RMS gas peculiar velocities $\sigma_\text{gas}$. 
This is not a property of the reconstruction method, but of the Universe.
In our simulated analysis, we assume that the gas bulk velocity is the same as that of the host dark matter halo. We therefore simply identify $v_\text{gas}$ as the velocity of the host halo's center of mass;
\item The RMS reconstructed velocities $\sigma_\text{rec}$;
\item Their correlation coefficient object by object
\beq
r \equiv 
\frac{\langle v_{\rm gas} v_{\rm rec} \rangle}{\sigma_\text{gas} \sigma_\text{rec}}.
\label{eq:r_coeff}
\eeq
\end{itemize}

Thus, for an unbiased inference of gas properties from kSZ, we need to quantify $\sigma_\text{rec}$ and $r$.
These will be our key performance metrics throughout the paper.

For an unbiased inference of the gas profile from kSZ, we also need to know $\sigma_{\rm gas}$.
We expect the bulk motion of the gas to more closely match the bulk motion of the dark matter halo, rather than that of individual galaxies within it.
We thus simply assume that $\sigma_{\rm gas} \approx \sigma_{\rm halos}$.
As shown in \cite{Sheth_2001}, $\sigma_{\rm halos}$ would depend on the cosmology, the host halo mass and the environment.
However, we find that $\sigma_{\rm lin}$ is actually a good approximation, as it only differs in $3.5\%$ with respect to $\sigma_{\rm halos}$ as shown in  Fig.~\ref{fig:vrec_vcent_vsat_1D_histogram}.
This translates to an error of $\sim 3.5\%$ on the distribution of the missing baryons, which is an error of $\sim 0.5\%$ on the total matter distribution, since baryons are 16$\%$ of the total matter density in the Universe \cite{Planck2020}.
This uncertainty of $0.5\%$ on the matter distribution is appropriate to model future galaxy lensing measurements from e.g., Rubin LSST, which are going to have percent precision.
For an even more accurate estimation, and for a consistent joint analysis with lensing, one may wish to incorporate an N-body simulation emulator to predict the cosmology dependence of $\sigma_\text{halos}$.

Furthermore, schematically, the noise on a kSZ estimator $\epsilon$ is
\beq
\bal
n_\epsilon 
&\propto \delta T_\text{noise}\  v_\text{rec} / \sqrt{N_\text{gal}} \\
&\propto \sigma_{\rm noise} \sigma_\text{rec} / \sqrt{N_\text{gal}},
\eal
\eeq
where $N_\text{gal}$ is the total number of galaxies, $\sigma_{\rm noise}$ is the noise standard deviation in the temperature map (e.g. detector noise, primary CMB, foregrounds, atmosphere, etc.), and we have assumed the galaxy sample to be shot noise limited\footnote{The scaling of the noise as $1/\sqrt{N_\text{gal}}$ breaks down once the galaxy sample is dense enough that several galaxies appear inside the same CMB map pixel, or more generally within patches with correlated noise. While this will be the case for Rubin galaxies, this toy example is still a valuable picture to have in mind.}.

Thus the signal-to-noise ratio (SNR) is simply proportional to the correlation coefficient, multiplied by the square root of the number of galaxies:
\beq
\text{SNR}
\propto r  \sqrt{N_\text{gal}}.
\eeq
As a result, the correlation coefficient $r$ is the actual figure of merit for the velocity reconstruction, which we seek to maximize.
Furthermore, a lower quality velocity reconstruction (i.e., lower $r$), can visibly be compensated by a larger sample size to yield the same kSZ SNR.

\begin{figure}[H]
     \centering
     \begin{subfigure}[b]{0.48\textwidth}
         \centering
         \includegraphics[width=0.99\textwidth]{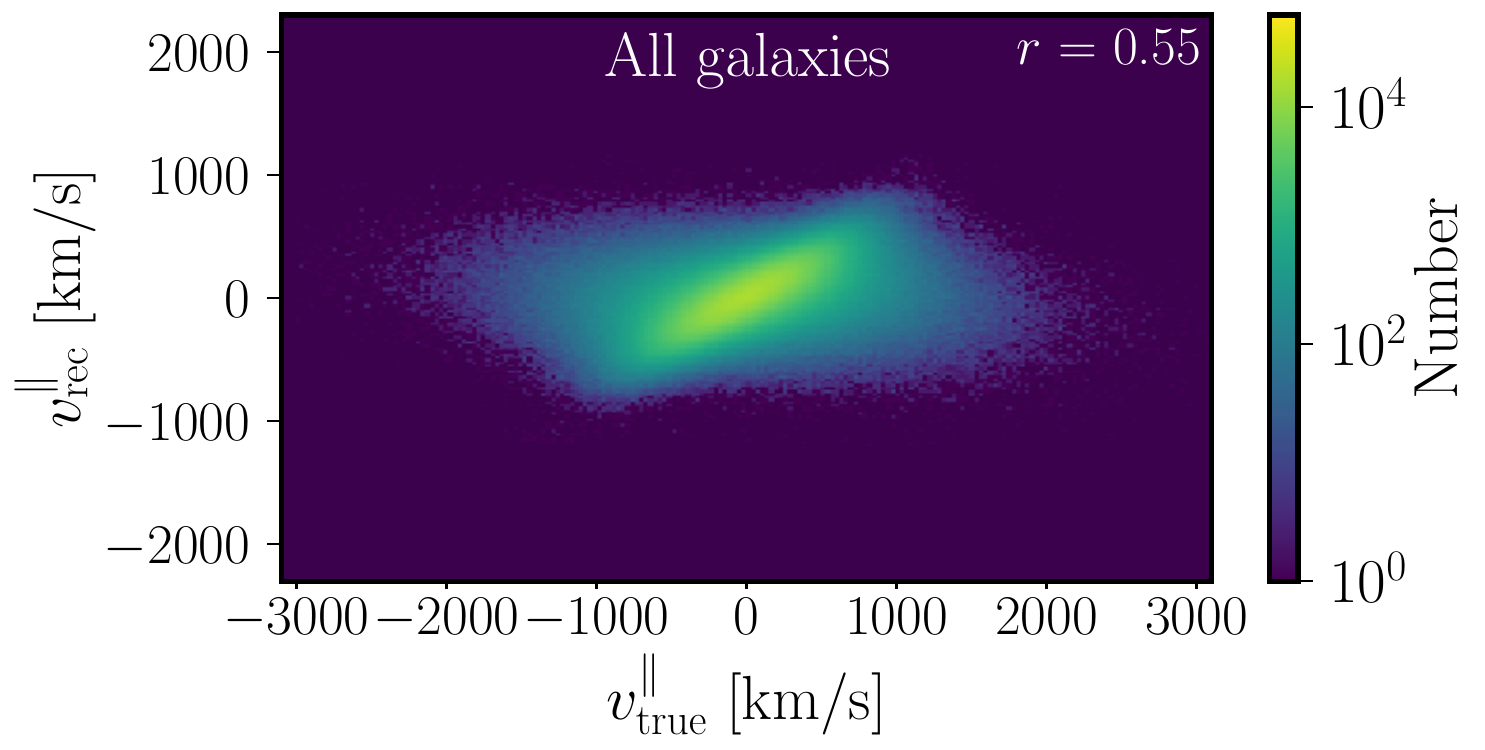}
     \end{subfigure}
     \hfill
     \begin{subfigure}[b]{0.48\textwidth}
         \centering
         \includegraphics[width=0.99\textwidth]{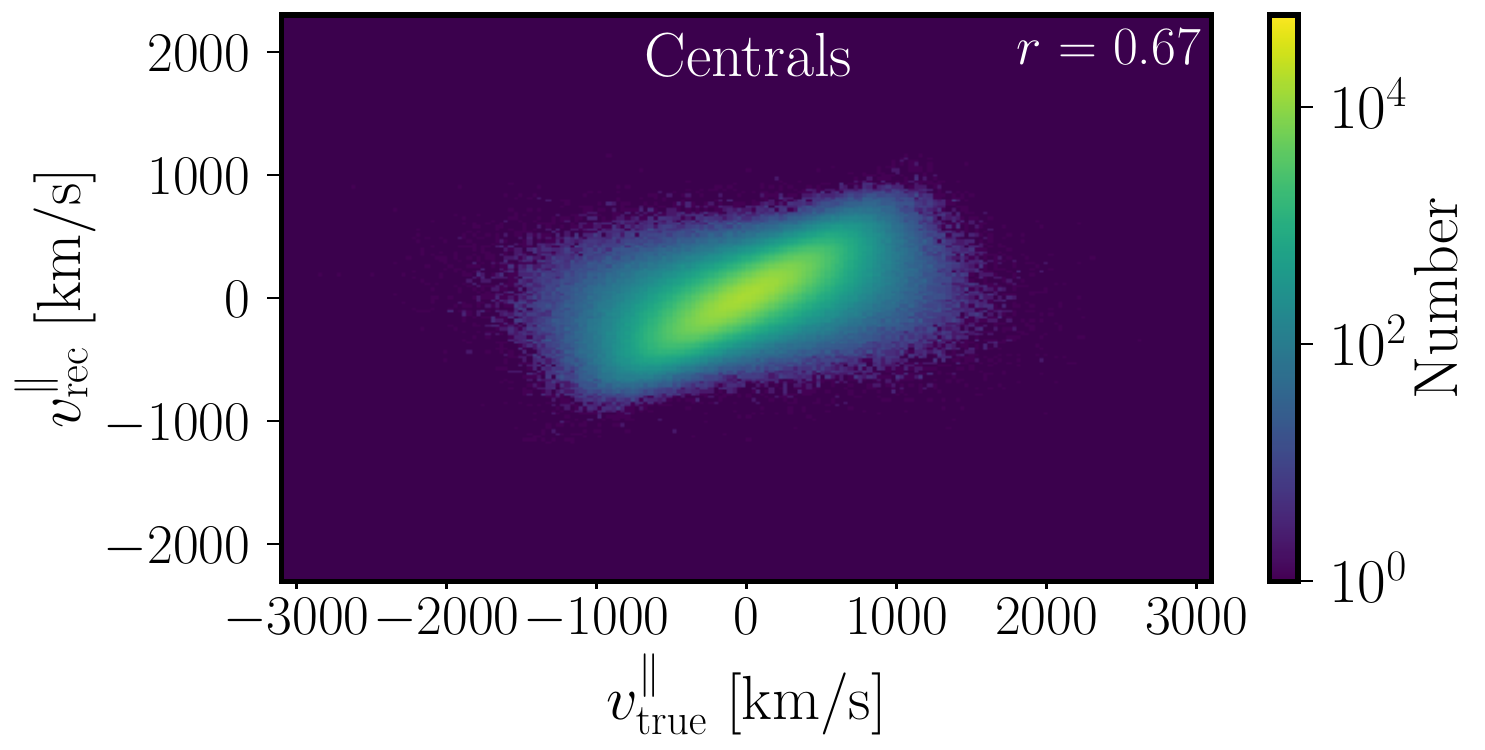}
     \end{subfigure}
     \hfill
     \begin{subfigure}[b]{0.48\textwidth}
         \centering
         \includegraphics[width=0.99\textwidth]{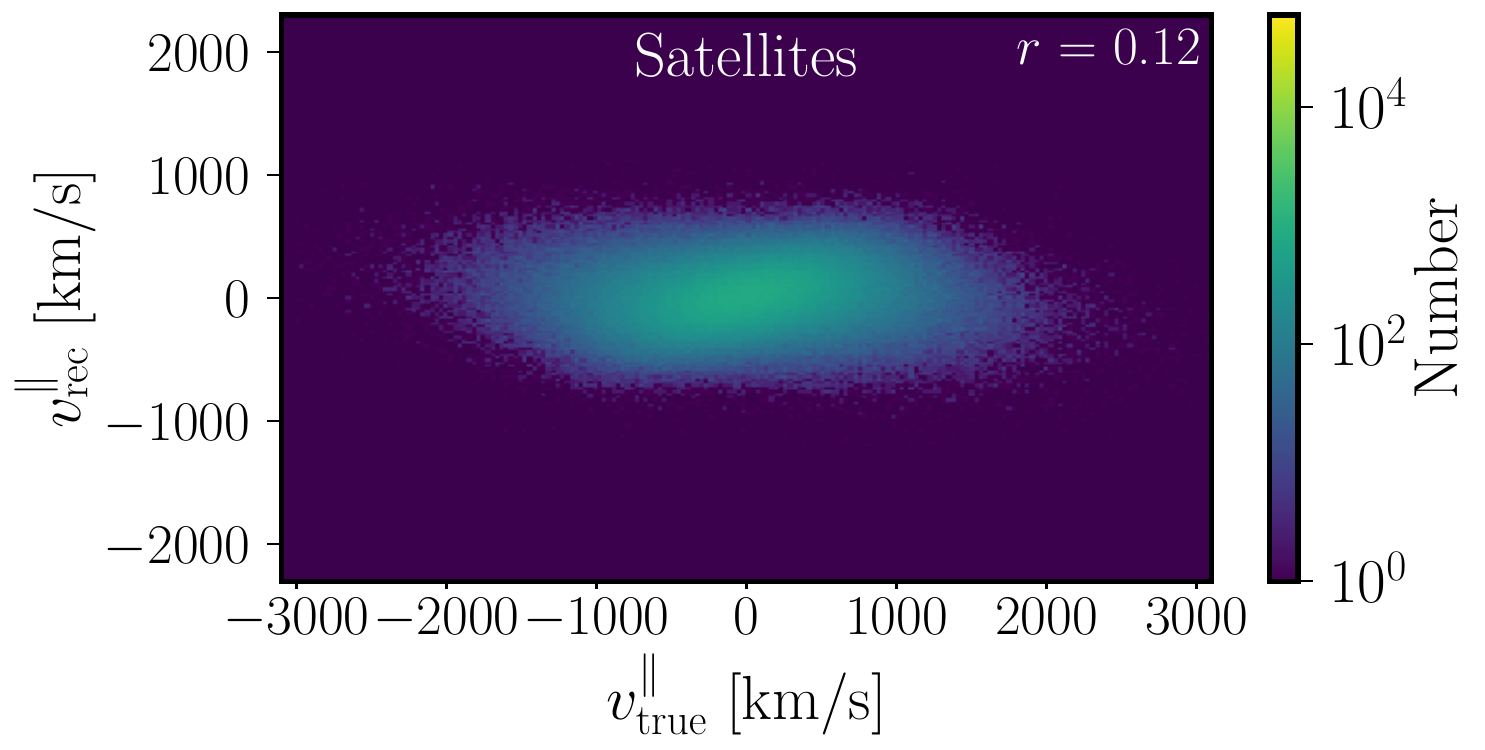}
     \end{subfigure}
     \caption{
        2D histograms of reconstructed v/s true galaxy velocities along the LOS for the fiducial \textsc{AbacusSummit} box.
        The distributions show a nearly-Gaussian core with positive correlation (center yellow ellipse),
        and tails at larger velocities where the correlation is null or even negative.
        These tails are due to nonlinear motion, including virial motion, which also cause unmodeled redshift-space distortions (Fingers-of-God).
        As expected, this tail is more important for satellites (bottom) than centrals (middle); the overall galaxy sample (top) being simply a mixture of the two.
        See App.~\ref{app:across_LOS} for the velocity components across the LOS.
        }
    \label{fig:vrec_vcent_vsat_histogram_los}
\end{figure}

For the fiducial values of our cosmological parameters, DESI LRG galaxy sample and smoothing scale from Table.~\ref{tab:fiducial_parameters}, we find:
\beq
\left\{
\bal
\sigma_{\text{gal}} = 369\ \text{km/s}\\
\sigma_{\rm rec} = 201\ \text{km/s}\\
\eal
\right.
,
\eeq
for both across and along the LOS.

\begin{figure}[H]
    \centering
    \includegraphics[width=0.48\textwidth]{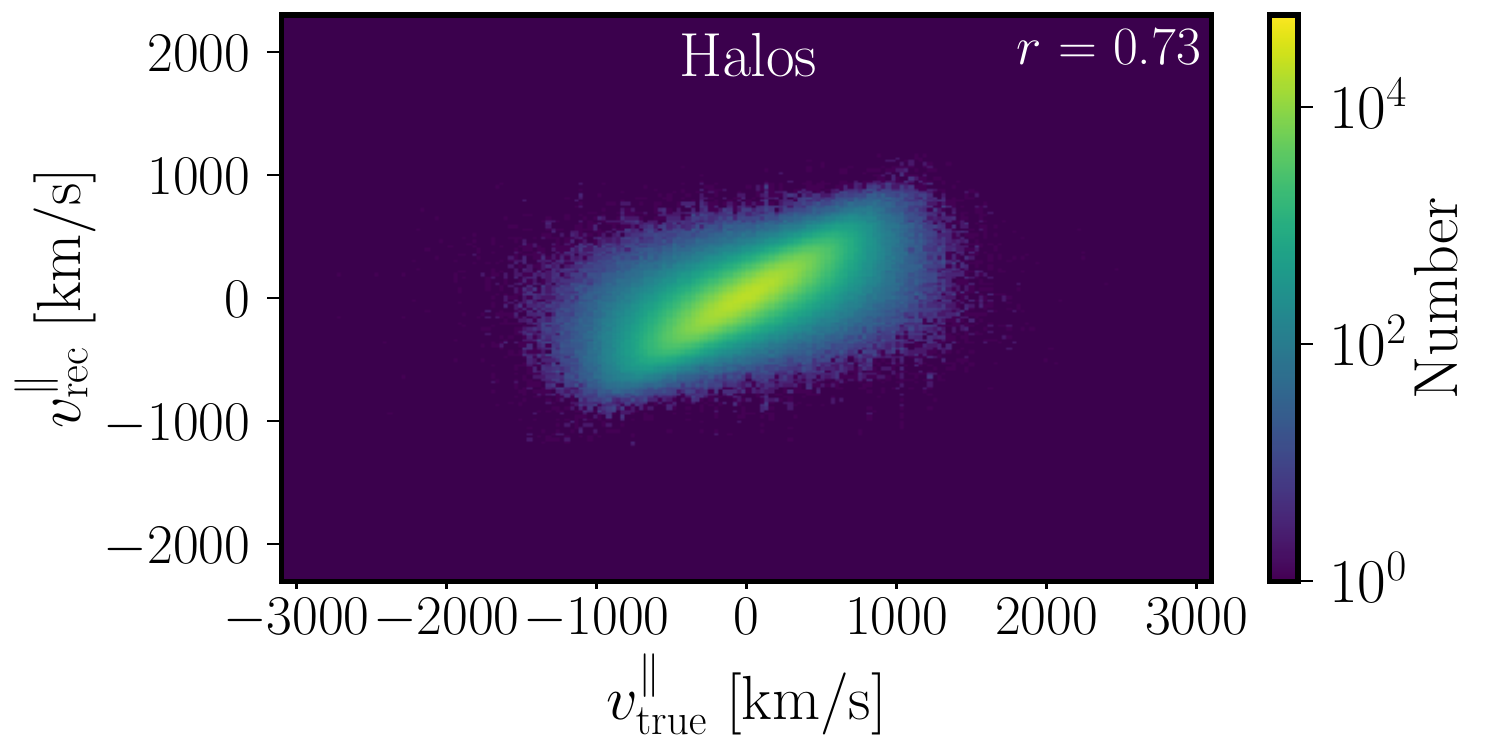}
    \caption{
        For the purpose of kSZ or moving lens stacking, the velocities we are actually trying to reconstruct are those of the halos (this figure), rather than those of the individual galaxies (Fig.~\ref{fig:vrec_vcent_vsat_histogram_los}).
        Fortunately, the reconstructed velocities are closer to the halo velocities than the galaxy velocities, since they are less affected by the non-linear evolution of the matter density field and the Finger-of-God effects.
        See App.~\ref{app:across_LOS} for the component across the LOS.
        }
        \label{fig:vrec_vhalos_2D_histogram_los}
\end{figure}

In agreement with Fig.~\ref{fig:cubic_box_visualization}, the linear approximation and the smoothing scale result in reconstructed speeds lower than the truth.
Interestingly, the RMS reconstructed velocity is the same along and across the LOS, suggesting that the Kaiser RSD is properly accounted for in our analysis, and that the FoG RSD displacements are mostly below our smoothing scale.

%%%%%%%%%%%%%%%%%%%%%%%%%%%%%%%%%%%%%%%%%%%%%%%%%%%%%%%

\subsection{Reconstructing halo rather than galaxy velocities}
\label{sec:Halo_vs_galaxies}

For kSZ measurements, the peculiar velocity we seek to estimate is that of the gaseous halo; for moving lens, it is the total halo velocity (dark matter plus gas).
As a result, the performance metrics above should really be applied to compare the reconstructed velocities to the true halo velocities, not the true galaxy velocities.
This distinction is significant, as shown in Figs.~\ref{fig:vrec_vcent_vsat_histogram_los} and~\ref{fig:vrec_vhalos_2D_histogram_los}.

The orbital motion of satellite galaxies within halos lead to large virial speeds.
Indeed, Fig.~\ref{fig:vrec_vcent_vsat_histogram_los}  shows that the  galaxy velocities are much better reconstructed for the centrals (middle panel, $r=0.67$) than the satellites (bottom panel, $r=0.12$).
This is expected due to the highly nonlinear virial motion of satellites, which is not accounted for in our linear reconstruction.
In the satellite population, the LOS reconstruction estimate has the wrong sign more often than for centrals.
We interpret this as galaxies whose virial LOS velocity is high and therefore, in redshift space these may be displaced to the opposite side where the central halo is.
Therefore, when using the velocity reconstruction method, their estimated velocity along the LOS points to the original position in real space.
In Fig.~\ref{fig:vrec_vcent_vsat_histogram_across} we show the case of velocities across the LOS, and, as expected, we find a better overall performance of the reconstruction method due to the absence of RSD.

However, for the purpose of kSZ or moving lens, not reconstructing the virial motion of satellites is actually an advantage: these virial motions are not shared by the dark matter or gaseous halos of interest.
Indeed, the correlation coefficient is higher with halo velocities ($r=0.73$, Fig.~\ref{fig:vrec_vhalos_2D_histogram_los}) than for galaxy velocities ($r=0.55$, Fig.~\ref{fig:vrec_vcent_vsat_histogram_los}).
See Fig. \ref{fig:vrec_vhalos_2D_histogram_across} for the velocity component across the LOS.
The performance of the linear velocity reconstruction would thus be much underestimated if we used galaxy velocities as the truth, rather than halo velocities.
Incidentally, the true 1D RMS velocity of our halos is $\sigma_{\rm halos} = 314$ km/s, lower than for galaxy velocities ($\sigma_{\rm gal} = 369$ km/s), and roughly exceeding the matter field ($\sigma_{\rm lin} = 303$ km/s).

%%%%%%%%%%%%%%%%%%%%%%%%%%%%%%%%%%%%%%%%%%%%%%%%%%%%%%%
%%%%%%%%%%%%%%%%%%%%%%%%%%%%%%%%%%%%%%%%%%%%%%%%%%%%%%%

\section{\boldmath Performance impact of RSD, photo-$Z$, galaxy properties, cosmology \& hybrid spectro-photometry}
\label{sec:results}

In this section we study the effect of various crucial parameters on the performance of the linear reconstruction method, using the correlation coefficient $r$ between reconstructed velocities and true halo velocities, and the standard deviation of the reconstructed velocities.

%%%%%%%%%%%%%%%%%%%%%%%%%%%%%%%%%%%%%%%%%%%%%%%%%%%%%%%

\subsection{\boldmath Photo-$z$ uncertainties: compensated by the increased sample size}
\label{sec:photo_z}

While spectroscopic redshifts are substantially more precise than photometric redshifts, photometric samples from e.g., the Rubin Observatory LSST \cite{lsstsciencecollaboration2009lsst}, Euclid \cite{Amendola_2018} or Roman \cite{Spergel_2015} can be orders of magnitude larger in size than the largest spectroscopic surveys, e.g., DESI \cite{Aghamousa2016} or PFS \cite{Takada_2014}.
As shown above, any degradation on the velocity reconstruction correlation coefficient from photometric redshift uncertainties will reduce the SNR of kSZ or moving lens as SNR $\propto r$ (see \cite{chan2023reconstructing} for a similar transverse reconstruction, but for the BAO signal in the presence of photo-z uncertainties).
However, in the shot noise regime, the SNR also grows as the square root of the total number of galaxies, SNR $\propto \sqrt{N_\text{gal}}$.
As a result of this trade-off, a photometric sample will lead to a larger kSZ or moving lens SNR is it has a higher $r\ \sqrt{N_\text{gal}}$.
To simulate the effect of photo-$z$ errors, we add Gaussian noise to the individual redshifts of the mock galaxies, 
\begin{equation}
    z_{\rm photo} = z + \mathcal{N}(0, \sigma_z)
\end{equation}
with $\sigma_z/(1 + z)$ in the range $[0.00 - 0.14]$.
When these new ``photo-$z$'' are converted to 3D positions as part of the reconstruction algorithm, they thus introduce a shift in LOS comoving positions as
\beq
\delta \chi = \frac{c}{H} dz.
\eeq
\begin{figure}[H]
    \centering
    \includegraphics[width=0.48\textwidth]{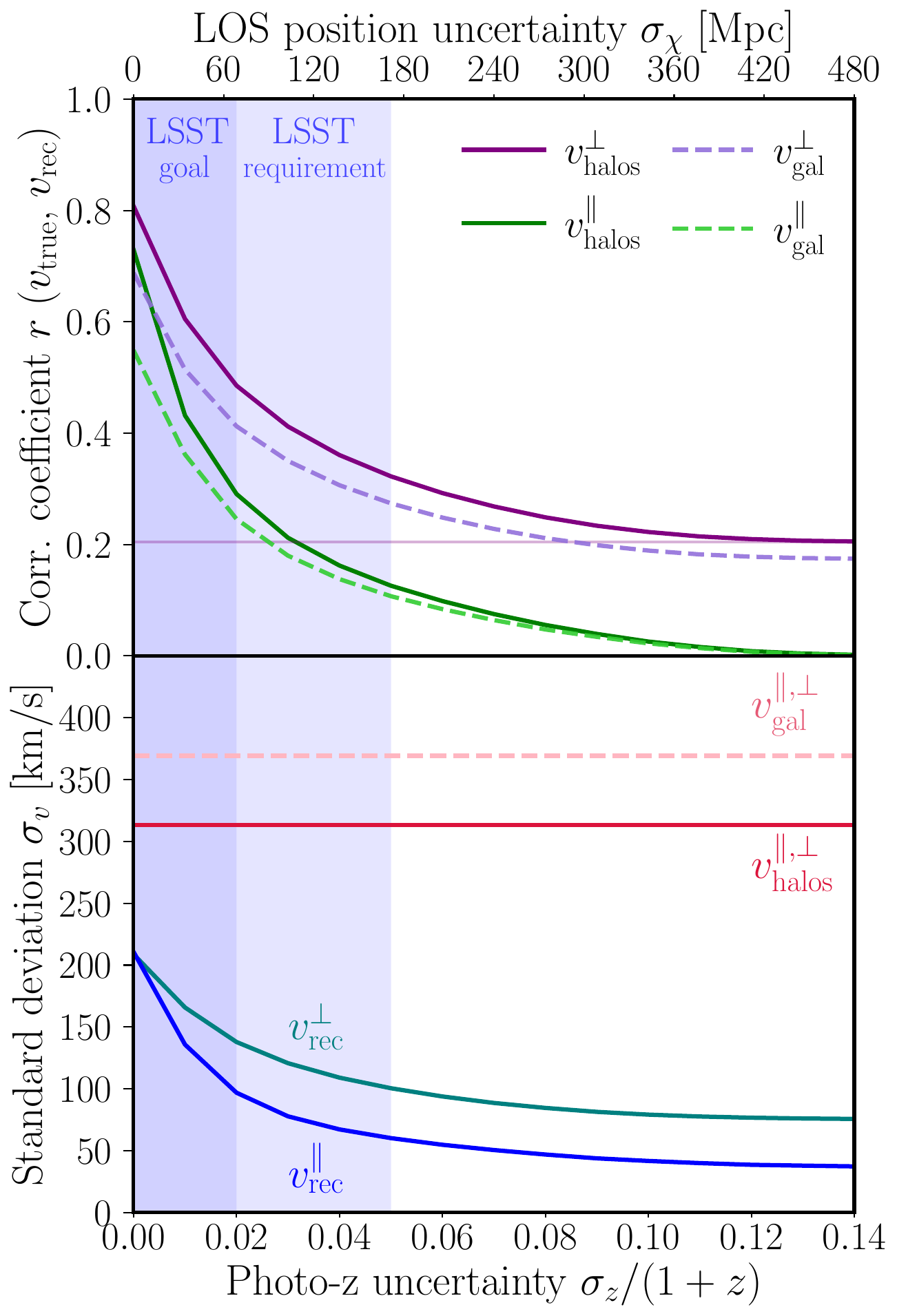}
    \caption{
    Effect of photo-$z$ uncertainties on the velocity reconstruction.
    \textit{Upper panel}:
    As the photo-$z$ uncertainty increases,
    the correlation coefficient $r = \frac{\langle v_{\text{halos}} v_{\text{rec}} \rangle}{\sigma^{v}_{\text{halos}} \sigma^{v}_{\text{rec}}}$
    decreases faster for the velocity component along the LOS than across.
    The non-zero correlation coefficient across the LOS in the limit of infinite photo-$z$ uncertainty (i.e. no redshift information) may be an artefact of the finite size and periodicity of the simulation box.
    The purple vertical bands corresponds to the LSST goal and requirement of the photo-$z$ root-mean-square scatter \cite{LSST_SB_2009}.
    \textit{Bottom panel}:
    As expected, the standard deviation of $v_{\rm halos}$ does not get affected by the photo-$z$ errors.
    On the other hand, in the absence of photo-$z$ errors, $v^{\|}_{\rm rec}$ and $v^{\perp}_{\rm rec}$ are the same despite redshift space distortions. 
    As the photo-$z$ uncertainty increases, the variance of the reconstructed density field is lower because photo-$z$ errors effectively smooth the observed galaxy number density field.
    }
    \label{fig:photo_z}
\end{figure} 

Fig. \ref{fig:photo_z} shows the degradation in $r$ and reduction in $\sigma_v$ as the photo-$z$ uncertainty increases.
Unsurprisingly, photo-$z$ errors are more damaging to the component of the velocity along the LOS, rather than across.
Indeed, photo-$z$ errors smooth out the LOS Fourier modes on scales smaller than $\sigma_\chi = H \sigma_z / c$, thus reducing the information available to the LOS velocity reconstruction.

As shown in Fig. \ref{fig:photo_z}, for typical LSST photo-$z$ uncertainties ($\sigma_z/(1+z) = 0.02$ or $0.05$ for ``goal'' and ``requirement'' \cite{LSST_SB_2009}), the corresponding smoothing scale is $\sigma_{\chi}= 67$ Mpc or $168$ Mpc, which are comparable to the typical coherence length of the true velocity field. 
As a result, an order unity degradation in $r$ occurs.
For LSST ``goal'' photo-$z$ uncertainties, $r$ is degraded by a factor 1.6 along and 2.5 across the LOS.
Again, the corresponding degradation in SNR can be easily compensated by the increased number of galaxies.

Eventually, $r$ goes to zero along the LOS for large enough photo-$z$ errors, as expected.
Intriguingly, $r$ asymptotes to a non-zero value at large photo-$z$ errors.
This would be very helpful for measurements of the moving lens effect.
In a reasonable redshift range, we could still reconstruct transverse velocities of 2D surveys.
However, this result may also be an artefact of our finite-sized box, 2 Gpc/$h$ on the side, and the periodic boundary conditions.
We leave this investigation for future work. 
Indeed, velocities across the LOS are sourced by differences in density from side to side in the box. While the photo-$z$ error reshuffles the LOS galaxy positions, it does not affect their positions across the LOS.

%%%%%%%%%%%%%%%%%%%%%%%%%%%%%%%%%%%%%%%%%%%%%%%%%%%%%%%

\subsection{Insensitivity to galaxy number density}
\label{sec:number_density}

The galaxy number density, i.e. the number of galaxies per unit cosmic volume, depends on the survey depths, the galaxy type, and the redshift of interest.
Spectroscopic surveys like DESI reach e.g., $\sim 10^{-3}$ ($h^{-1}$Mpc)$^{-3}$ for LRG galaxies \cite{Yuan2023}.
How does this parameter affect the velocity reconstruction?
We answer this question by randomly downsampling our mock galaxy sample, to simulate samples with lower number densities.

Interestingly, Fig.~\ref{fig:density} shows that the correlation coefficient and standard deviation of the reconstructed velocities are fairly constant over a substantial range of number densities, down to $\sim 5 \cdot 10^{-5}$ Mpc$^{-3}$.
This includes DESI LRGs and QSOs, BOSS LRGs, and even marginally SDSS LRGs.
The performance for DESI QSOs is degraded by a factor 1.4 across the LOS.
Thus, for our purposes, current and future experiments already saturate the performance in terms of number density.
Further increases in number density will not improve kSZ and moving lens SNR using this reconstruction method.
However, increases in total number of galaxies, i.e. in survey volume, will. 
This may be useful input for future survey planning.

Intuitively, increasing the galaxy number density reduces the galaxy shot noise, making smaller and smaller scales accessible for the velocity reconstruction.
In the upper horizontal axis, we thus give the conversion between number density and $k_{\rm shot}$, the Fourier scale at which shot noise starts to dominate over galaxy clustering, i.e.:

\begin{figure}[H]
    \centering
    \includegraphics[width=0.48\textwidth]{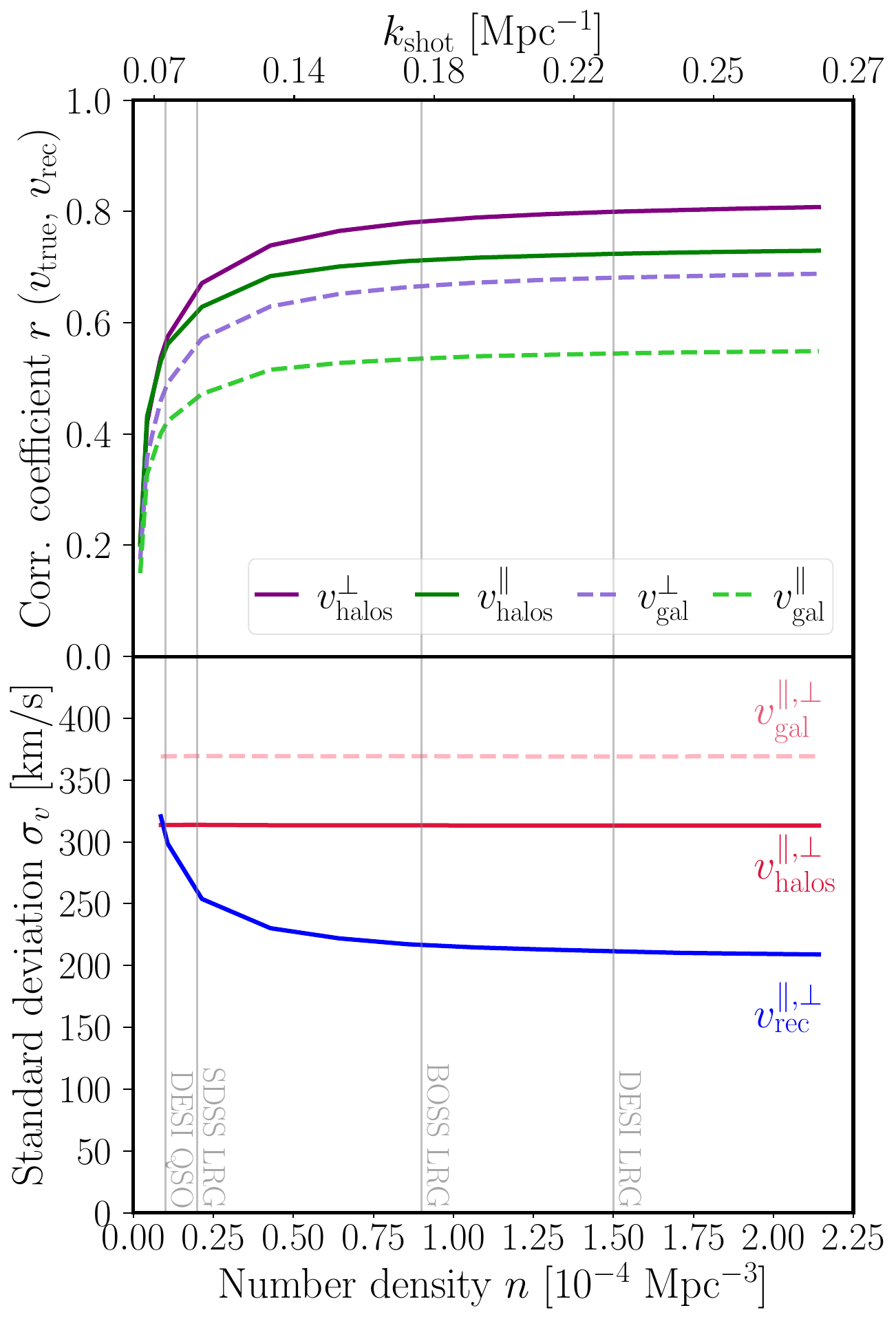}
    \caption{
    Impact of the galaxy number density on the velocity reconstruction performance.
    The vertical gray lines correspond to the mean co-moving number densities of relevant LRG and QSO samples \cite{Eisenstein2005, Reid2016, Yuan2023, rocher2023desi}.
    \textit{Upper panel}:
    As the number density increases the correlation coefficient saturates, despite the reduction in shot noise. This indicates that our reconstruction algorithm does not leverage small-scale information, in part due to the linear approximation, and in part due to the fixed smoothing scale (14.8 Mpc). The top abscissa indicates the effective smoothing scale due to shot noise introduced in Eq.~\eqref{eq:shot_noise_scale}.
    }
    \label{fig:density}
\end{figure} 
\begin{equation}
    b^2 P_{m}(k_{\rm shot}) = \frac{1}{n}
\label{eq:shot_noise_scale}
\end{equation}
where $P_{m}(k)$ is the linear power spectrum and $b$ the linear bias.

Thus, the value $\overline{n} \sim 5 \cdot 10^{-5}$ Mpc$^{-3}$ where the performance saturates corresponds to scales of $k_\text{shot} \sim 0.1/$Mpc, or $\sim 60$ Mpc, larger than our smoothing scale of $14.8$ Mpc.
This confirms that this saturation is not due to our smoothing scale, but really a feature of the linear reconstruction.

%%%%%%%%%%%%%%%%%%%%%%%%%%%%%%%%%%%%%%%%%%%%%%%%%%%%%%%

%
\begin{figure}[H]
    \centering
    \includegraphics[width=0.48\textwidth]{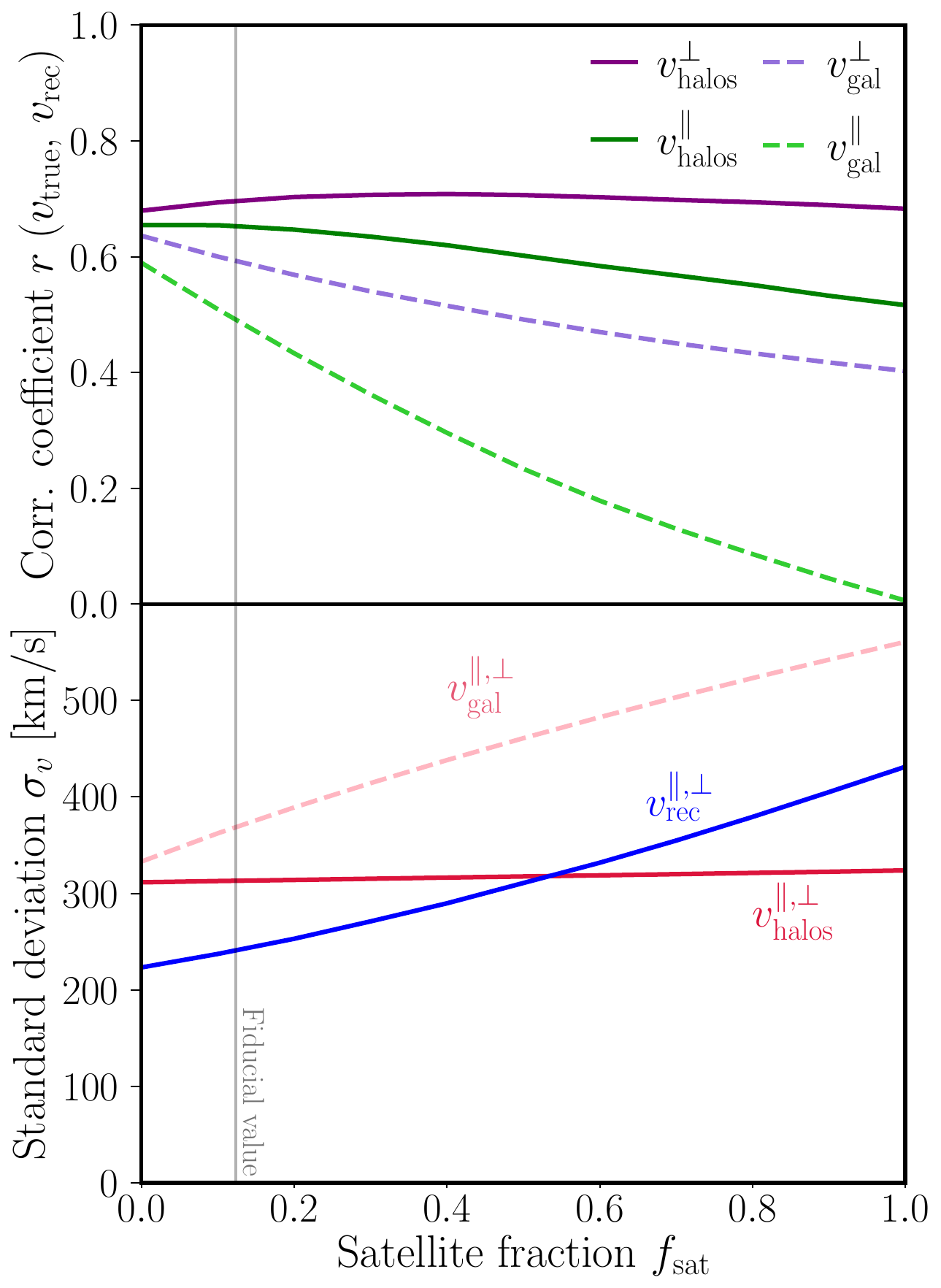}
    \caption{
    Impact of the satellite fraction on the velocity reconstruction.
    The vertical line corresponds to the fiducial value $f_{\rm sat} = 0.12$.
    \textit{Upper panel}:
    Changing the satellite fraction does not affect significantly the correlation coefficient measured on $v^{\perp}_{\rm halos}$, but does impact $v^{\|}_{\rm halos}$.
    For the galaxy sample, the satellite fraction affects both $v^{\perp}_{\rm gal}$ (due to non-linearities) and more drastically $v^{\|}_{\rm gal}$ (due to non-linearities and RSD).
    \textit{Bottom panel}:
    Satellite galaxies, as shown in Fig. \ref{fig:vrec_vcent_vsat_1D_histogram}, have a larger distribution of velocities due to their non-linear motions. 
    When their fraction is increased and the number density remains fixed to $\overline{n} = 8 \cdot 10^{-5}$ Mpc$^{-3}$, the standard deviations of $v^{\|, \perp}_{\rm rec}$ and $v^{\|, \perp}_{\rm gal}$ grow substantially, while the halo velocities $v^{\|, \perp}_{\rm halos}$ are almost unchanged.
    }
    \label{fig:sat_fraction}
\end{figure} 

\subsection{Robustness to satellite fraction variations}
\label{sec:sat_fraction}

Different tracer samples, such as LRGs, ELGs and QSOs, inhabit different halos and in different ways. 
For instance, the satellite fraction in ELGs is much higher than in LRGs \cite{rocher2023desi, yuan2023desi, yuan2023unraveling}.
As described above, satellite galaxies have larger virial motions, which are not relevant for kSZ and moving lens.
For example, satellite LRGs cause larger FoG, thus smoothing out the observed number density in redshift space and reducing the information available for the velocity reconstruction.
How much does this affect the velocity reconstruction?

To study the response of the velocity reconstruction to a change in the satellite fraction $f_{\rm sat}$, we randomly downsample the centrals or satellite galaxies.
However, we want to vary $f_{\rm sat}$ at fixed number density, for a fair comparison.
Randomly downsampling the whole catalog to $\overline{n} = 8 \cdot 10^{-5}$ Mpc$^{-3}$ allows us to keep the number density constant while spanning values of $f_{\rm sat}$ in $[0.0-1.0]$.
As shown in Sec.~\ref{sec:number_density}, this reduction in the overall number density should not substantially affect our analysis.

In Fig.~\ref{fig:sat_fraction} we show the performance of both the correlation coefficient $r$ and the standard deviations $\sigma_v$ as we change the satellite fraction.
In this case, we find that reconstructing the velocities with respect to $v_{\rm halos}$ rather than $v_{\rm gal}$ is determinant.

Indeed, when comparing reconstructed velocities to galaxy velocities, an increased satellite fraction degrades the correlation coefficient (top panel).
The effect is more dramatic along the LOS again, as expected due to the increased FoG, leading to $r \sim 0$ when $f_{\rm sat} \sim 1$.
On the other hand, when comparing $v_{\rm rec}$ to $v_{\rm halos}$, we find a better match.
The reconstruction correlation coefficient across (resp. along) the LOS is unaffected (resp. only mildly reduced) as the satellite fraction increases.
Focusing on the lower panel of Fig.~\ref{fig:sat_fraction}, we find the RMS halo velocities to be independent of satellite fraction, whereas the galaxy velocities increase, as expected due to the large fraction of objects with virial motions.

Intriguingly, the RMS reconstructed velocities do increase with satellite fraction, in the same measure for $v^{\|}$ and $v^{\perp}$.
This is explained by the way the HOD is constructed: 
when selecting a higher fraction of satellite galaxies, we are selecting more massive halos.
While these more massive halos have roughly the same true velocities (red curve in Fig.~\ref{fig:sat_fraction}), they also have a larger clustering bias $b_{\rm eff}$.
Since we do not account for this variation in clustering bias in the reconstruction, the matter density field we infer from the galaxy catalog is suriously amplified by a multiplicative factor $b_\text{eff}/b_\text{fiducial}$.
However, this factor boosts the standard deviation equally across and along the LOS, and does not affect the correlation coefficient.
As a caveat, we state that down-sampling randomly is not quite the same as e.g. going down to a higher apparent magnitude in a realistic scenario.
This simple exploration aims to gain intuition and does not reflective the true behaviour of real samples in the universe.

%%%%%%%%%%%%%%%%%%%%%%%%%%%%%%%%%%%%%%%%%%%%%%%%%%%%%%%

\subsection{Optimal choice of smoothing scale}
\label{sec:smoothing_scale}

\begin{figure}[H]
    \centering
    \includegraphics[width=0.48\textwidth]{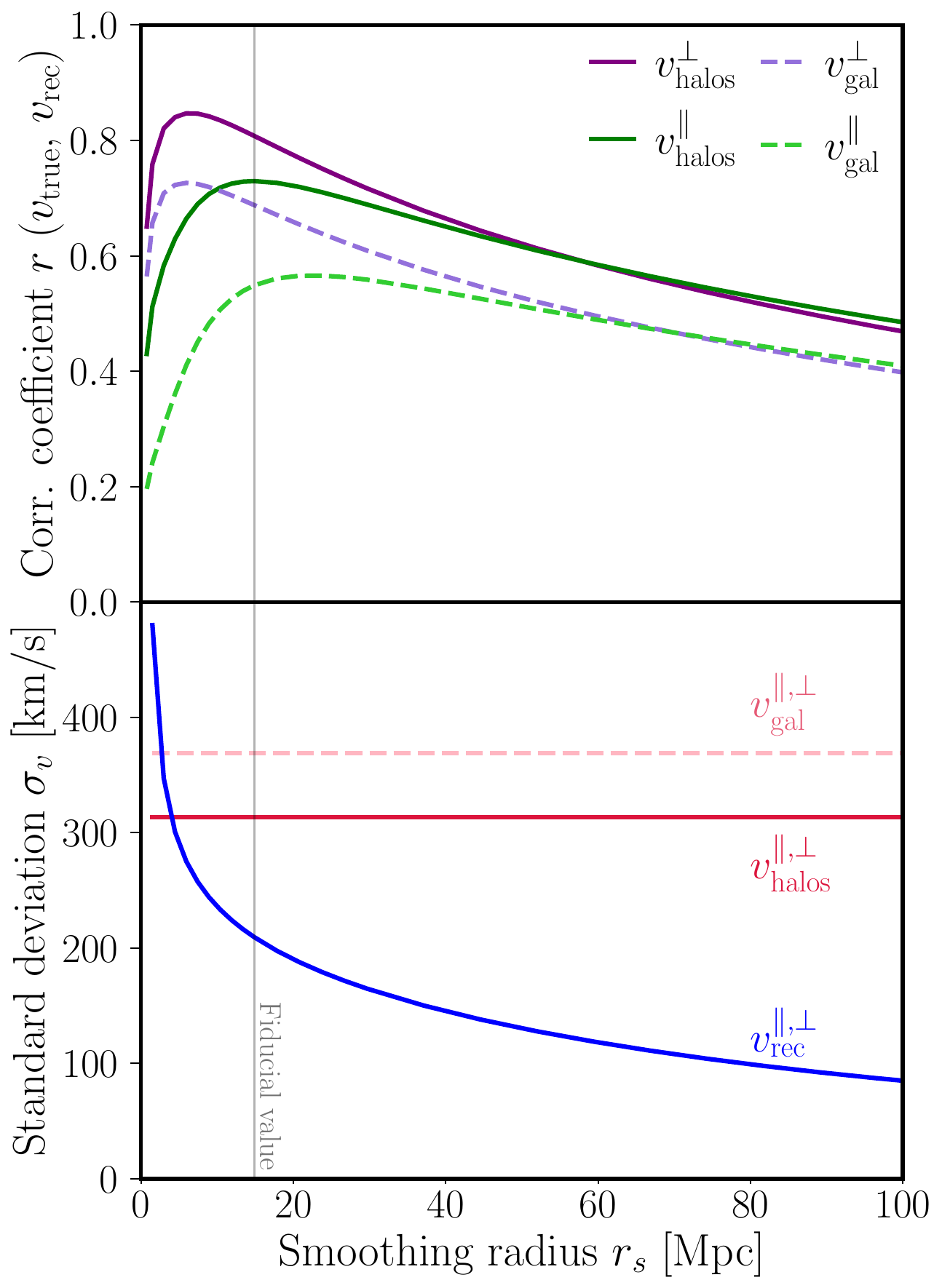}
    \caption{
    \textit{Upper panel}:
    For our fiducial galaxy number density, the correlation coefficients between reconstructed and halo velocities have an optimal smoothing radius $r^{\perp}_s = 5.9$ Mpc and $r^{\|}_s = 14.8$ Mpc. 
    We see a similar result in for $v_{\rm gal}$. 
    Smaller smoothing scales lead to the inclusion of nonlinear and shot noise dominated modes,
    while larger smoothing scales remove some of the useful linear modes.
    As the smoothing scale increases to become larger than the linear Kaiser RSD, the reconstruction performance becomes identical along and across the LOS.
    \textit{Bottom panel}:
    For a smaller smoothing radius, the shot noise of galaxies contaminates the signal, and therefore, the reconstructed standard deviation is larger than the one from the simulations.
    For a larger smoothing radius, the standard deviation of the reconstructed velocities decrease because we erase relevant modes of the matter density field used in the reconstruction method. 
    }
    \label{fig:smoothing}
\end{figure}

One arbitrary parameter in our linear velocity reconstruction is the scale $r_s$ with which we smooth the galaxy number density field.
The goal of this smoothing is to downweight the small-scale Fourier modes which are shot noise dominated.
On the other hand, the smoothing effectively nulls Fourier modes which would otherwise contain velocity information.
The optimal value of $r_s$ is thus a trade-off.
In Fig.~\ref{fig:smoothing}, we assess the performance of the velocity reconstruction as a function of the smoothing scale around the fiducial value $r_s=14.8$ Mpc.

The correlation coefficients have a well-defined peak, both with respect to $v_{\rm gal}$ and $v_{\rm halos}$.
Across the LOS, the peak happens at $r^{\perp}_s = 5.9$ Mpc, while along the LOS, the peak happens at our fiducial value $r^{\|}_s = 14.8$ Mpc, as proved in \cite{Hadzhiyska2023} in Table 1.
For smoothing radii smaller than the peak values, shot noise dominates and contaminates the signal.
This is also visible in the large standard deviation of the reconstructed velocities on the bottom panel. 
In the other extreme case, larger smoothing radius smooth fundamental density modes, thus dramatically reducing this standard deviation.
Interestingly, the optimal smoothing scale along and across the LOS is different.
Being unaffected by the small scale power from FoG, the reconstruction across the LOS benefits from including smaller scales.

%%%%%%%%%%%%%%%%%%%%%%%%%%%%%%%%%%%%%%%%%%%%%%%%%%%%%%%

\subsection{Impact of incorrect cosmological parameters,
or the need to jointly fit for baryons and cosmology}
\label{sec:wrong_cosmo}

When performing velocity reconstruction on a real galaxy sample, the actual input is the catalog of angular positions and redshifts for each galaxy.
The reconstruction algorithm therefore takes cosmological parameters as input, in order to convert these angles and redshifts into 3D comoving positions.
This conversion requires assuming a value for the Hubble parameter and the comoving angular diameter distance. 
Values for the galaxy bias and logarithmic growth rate $f$ are then assumed, in order to convert the galaxy number density field into the peculiar velocity field.
If these assumed values differ from the truth, the inferred peculiar velocities will be biased.
How large is this bias, and how does it propagate to measurements of the kSZ or moving lens effects?
In this section, we study this bias for the relevant parameters: $H$, $\Omega_m^0$ and $b$ on the velocity reconstruction.
The growth factor $f$ is mostly a derived parameter from $\Omega_m^0$, and thus doesn't need to be varied separately.
Notably, no assumptions on the amplitude of scalar fluctuations $A_S$, or its late time analog $\sigma_8$, are required in the linear velocity reconstruction, and these parameters therefore do not need to be varied.
As explained in detail in App. \ref{sec:algebra_wrong_params}, we vary these parameters in the reconstruction, away from the fiducial value that was used to generate the ``true'' velocities in the mock.
We show the resulting biases in Fig.~\ref{fig:cosmology} and discuss them below.
We find that the impact of using the wrong parameters on the correlation coefficient is sub percent for cosmological parameters within the Planck 5-$\sigma$ uncertainty. 

However, the velocity standard deviations can vary by several percent.
For the reconstructed velocities, this is no issue: the corresponding standard deviations can be measured from the data itself.
For the true velocity standard deviation (black lines in Fig.~\ref{fig:cosmology}), this can cause a $\sim 2\%$ bias within the Planck 3-$\sigma$ confidence region. 
This would then potentially bias our inference of the gas profile by $\sim 2\%$.

If this bias is too large to be acceptable, one way to avoid it is to include cosmology in a joint analysis by including in the likelihood the kSZ model as follows:
\begin{equation}
    \ln\mathcal{L}\left({\text{kSZ} | \text{cosmo, gas}}\right)
    \propto \frac{[\text{kSZ} - \text{gas model(cosmo, gas)}]^2}{\sigma_{\rm kSZ}^2} ,
\end{equation}
where the ``cosmo'' and ``gas'' parameters can now be consistently sampled.

Similarly, if the intent is to use kSZ measurements to constrain the baryonic effects in galaxy lensing, in order to improve the cosmological inference from lensing, a safe approach is to jointly vary cosmology and gas parameters in a consistent joint analysis:
\beq
\bal
    \ln\mathcal{L}\left({\rm kSZ, lensing | cosmo, gas}\right)
    &= \ln\mathcal{L}\left({\rm kSZ | cosmo, gas}\right)\\
    &+ \ln\mathcal{L}\left({\rm lensing | cosmo, gas}\right),\\ 
\eal
\eeq
where $\mathcal{L}({\rm kSZ, lensing | cosmo, gas})$ corresponds to the joint likelihood of the kSZ and galaxy-galaxy lensing (ggl) given a cosmology and a gas measurement.
We also assumed that the kSZ and ggl measurements are independent.
The likelihood of the galaxy lensing as:
\beq
\bal
    &\ln \mathcal{L}\left({\rm lensing | cosmo, gas}\right) \\
    &\propto \frac{[{\rm lensing} - \text{gas model(cosmo, gas)} - {\rm DM (cosmo)}]^2}{\sigma_{\rm lensing}^2} .
\eal
\eeq
Crucially, for the purpose of calibrating baryonic uncertainties in galaxy weak lensing, we are helped by the fact that the baryons only make up $16\%$ of the total matter density.
Thus, a $2\%$ bias on the gas profile from kSZ, due to incorrect cosmological parameters, is only a $0.16\times 2\% = 0.3\%$ uncertainty on the matter density.
This is most likely acceptable for upcoming galaxy lensing experiments like Rubin LSST, which aim to reach percent precision on the scales of interest.
In this case, marginalizing over cosmology is not required.

%%%%%%%%%%%%%%%%%%%%%%%%%%%%%%%%%%%%%%%%%%%%%%%%%%%%%%%

\subsubsection{Hubble parameter: isotropic re-scaling}

\begin{figure*}
    \includegraphics[width=1.0\textwidth]{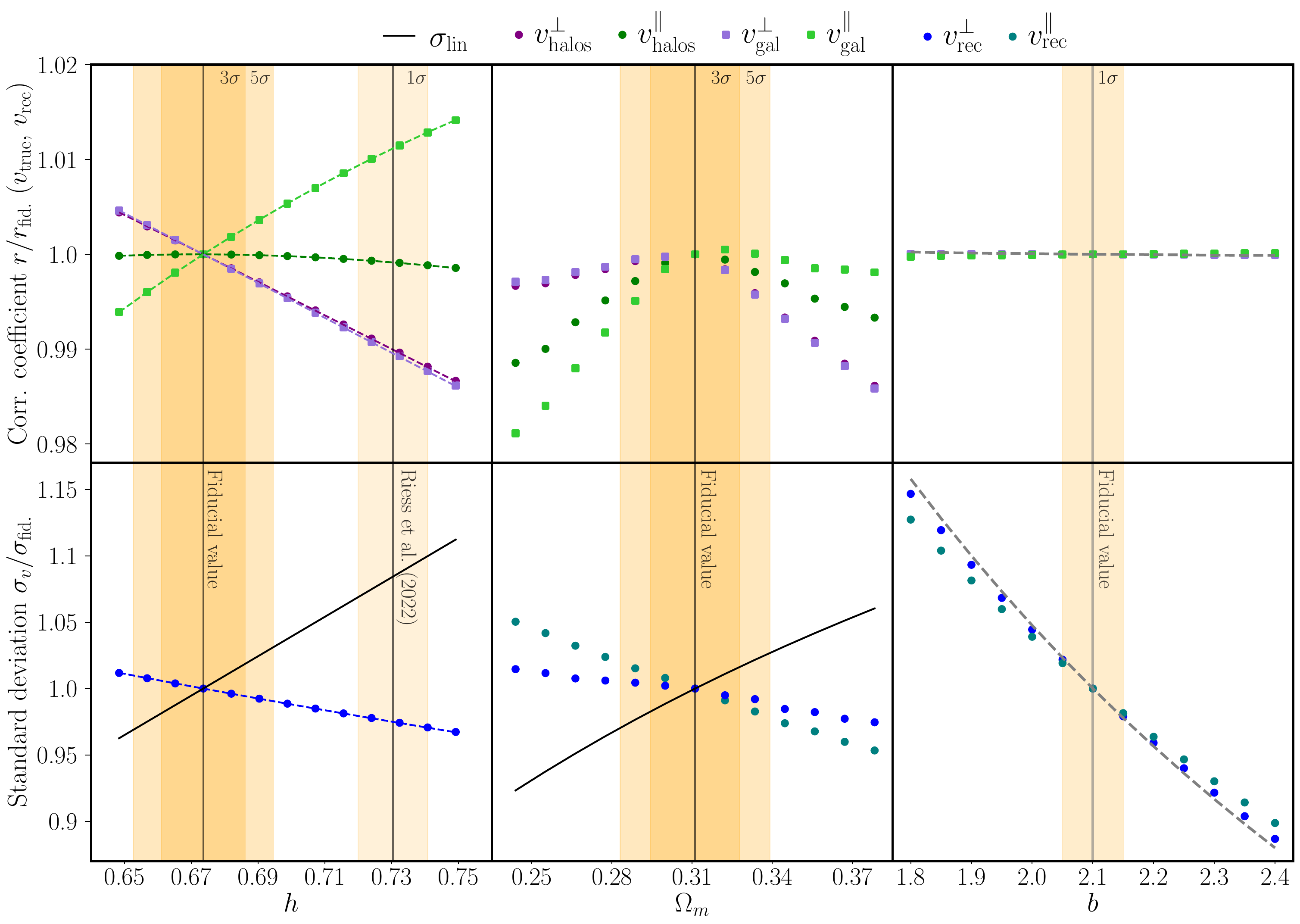}
    \caption{
    Variations in correlation coefficient $r/r_{\rm fid.}$ and standard deviation $\sigma_v/\sigma_{\rm fid.}$ when the reconstruction is performed assuming incorrect cosmological parameters, i.e. not the fiducial $h$, $\Omega_m$ and $b$ assumed to generate the mocks.
    For $h$ and $\Omega_m$, the yellow bands indicate the $\pm 3\sigma$ and $\pm 5\sigma$ uncertainties from Planck \cite{Planck2020} and the $\pm 1\sigma$ uncertainty on $h$ from \cite{Riess_2022}.
    For $b$, the yellow band corresponds to the $\pm 1\sigma$ uncertainty from DESI 1\% per cent survey \cite{yuan2023desi}.
    The black solid line reports the corresponding standard deviation of the matter density field from theory, refereed as $\sigma_{\rm lin}$ in Eq.~\ref{eq:th_vel_std_dev}.
    \textit{Upper panels: }
    Variations in cosmological parameters inside of Planck uncertainties only change the correlation coefficient at percent level.
    Uncertainties on the galaxy bias have no impact on $r$.
    \textit{Lower panels: }
    In the contrary, the standard deviation of the reconstructed velocities is most affected by errors in the assumed galaxy bias.
    Another visible feature is that errors in $\Omega_m$ and $b$ both introduce an anisotropic error in the reconstruction, leading to $\sigma_{v^{\|}_{\rm rec}} \neq \sigma_{v^{\perp}_{\rm rec}}$.
    For $h$, assuming an incorrect value only leads to an isotropic re-scaling of the box. 
    As derived in Sec.~\ref{sec:changing_h}, this is equivalent to a change in the smoothing scale $r^{\rm eff}_s =\frac{h_w}{h_t} r_s$ shown as dashed line in the $h$ plot.
    The result probes our algebraic expectations from Sec. \ref{sec:algebra_wrong_params}.
    }
    \label{fig:cosmology}
\end{figure*}

At the top left panel from Fig. \ref{fig:cosmology}, we find that as we increase the assumed $h$, even further than Planck's 5-$\sigma$ uncertainty from the fiducial value, the correlation coefficient does not change by more than 2\%.
As derived in App.~\ref{sec:changing_h}, changing $h$ simply corresponds to an isotropic re-scaling of the simulation box.
However, the smoothing scale is kept fixed (in comoving size), such that an increment/decrement in $h$ is exactly equivalent of fixing the box and smoothing larger/smaller scales.
We check this by predicting this effect based on the top panel from Fig. \ref{fig:smoothing} (dashed lines), which match our measurements (points).

For $h>h_{\rm fid.}$, both correlation coefficients calculated using $v^{\perp}_{\rm halos}$ and $v^{\perp}_{\rm gal}$ decrease because their peak happens, equivalently, at smaller smoothing radii.
For $v^{\|}_{\rm halos}$, the correlation is already in the peak of $v^{\|}_{\rm halos}$, and therefore, there is no substantial variation when changing $h$ on the plotted range.
For $v^{\|}_{\rm gal}$, the peak happens for larger smoothing radius, so the correlation increases with $h$.

This equivalence between changing $h$ and changing the smoothing scale is further confirmed in the bottom left panel.
Indeed, in this model, increasing the smoothing scale erases more of the velocity Fourier modes, leading to a reduced variance (dashed line), which matches the measured one (points).

Finally, in the lower panel we included $\sigma_{\rm lin}$ from Eq.~\ref{eq:th_vel_std_dev}. 
We find that it is impacted by the value of $h$, and therefore, reinforces the need of doing a joint likelihood analysis and a posterior marginalization over cosmology.

%%%%%%%%%%%%%%%%%%%%%%%%%%%%%%%%%%%%%%%%%%%%%%%%%%%%%%%

\subsubsection{Matter density: anisotropic re-scaling \& growth rate}

\begin{figure*}
    \includegraphics[width=0.9\textwidth]{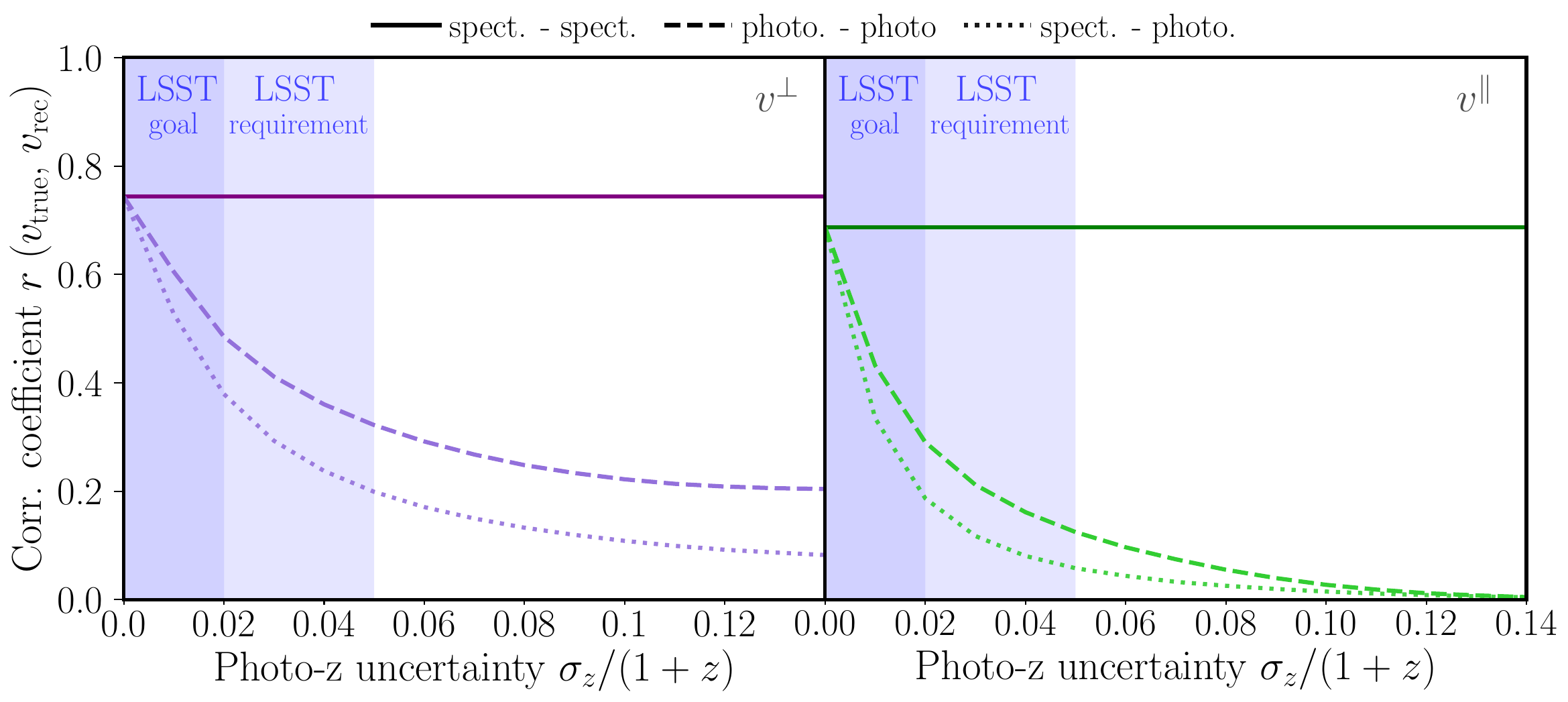}
    \caption{
    Consider a dense photometric sample and a sparse spectroscopic sample overlapping in 3D.
    If we wish to reconstruct the velocities of the photometric halos, several approaches are possible: each line labelled sample 1 - sample 2 indicates a method where the first step in the reconstruction (``'density estimation'') was performed with sample 1, and the the second step (``evaluating the velocities'') is done at the positions of sample 2 galaxies (see Sec.~\ref{sec:vel_rec}).
    Which one leads to the best correlation coefficient between reconstructed and true halo velocities, for the photometric sample?    
    Interestingly, the photometric - photometric reconstruction performs better than the spectroscopic - photometric reconstruction, despite the larger redshift uncertainty in the first step (``density estimation'').
    This surprising result can be understood intuitively, see App.~\ref{sec:hybrid_extra}.
    Any difference between the velocity component across (left) and along (right) the LOS is due to RSD.
    }
    \label{fig:hybrid_r}
\end{figure*}

Focusing on $\Omega_m^0$ (top central panel), the behavior is more complex.
Indeed, an incorrect value of $\Omega_m^0$ affects the Hubble parameter and the comoving angular diameter distance differently.
This in turn causes an isotropic rescaling of the simulation box (see  App.~\ref{sec:changing_omega}), though both directions along and across the LOS are increased for larger $\Omega_m^0$.
We find the bias on $r$ to be less than percent for Planck-like uncertainties, and a couple percent bias on the standard deviation.

We find:
\beq
\left\{
\bal
    & \frac{r_{\rm v^{\perp}_{\rm halos}}}{r_{\rm fid.}} 
    \sim \frac{r_{\rm v^{\perp}_{\rm gal}}}{r_{\rm fid.}}
    > \frac{r_{\rm v^{\|}_{\rm halos}}}{r_{\rm fid.}}
    > \frac{r_{\rm v^{\|}_{\rm gal}}}{r_{\rm fid.}}
    \text{ for } \Omega_m^w < \Omega_m^{\rm fid.} \\
    & \frac{r_{\rm v^{\perp}_{\rm halos}}}{r_{\rm fid.}} 
    \sim \frac{r_{\rm v^{\perp}_{\rm gal}}}{r_{\rm fid.}}
    < \frac{r_{\rm v^{\|}_{\rm halos}}}{r_{\rm fid.}}
    < \frac{r_{\rm v^{\|}_{\rm gal}}}{r_{\rm fid.}}
    \text{ for } \Omega_m^w > \Omega_m^{\rm fid.} \\
\eal
\right.
\eeq
with $r_{\rm fid.}$ the resulting correlation coefficient $r$ for our fiducial configuration.
The best correlation coefficient in most cases is when the correct value of $\Omega_m^0$ is assumed.
Indeed, any other value effectively leads to a re-weighting of the modes along the LOS, which worsen the reconstruction (see App.~\ref{sec:changing_omega}).

An increased matter density parameter reduces the standard deviations of the reconstructed velocity components along and across the LOS (bottom central panel).
This is expected again from the increased effective smoothing scale.
The different behavior along and across the LOS is a consequence of the differing multiplicative constants introduced in Eq.~\ref{eq:constants_Omega_m} and the different smoothing scales.

Ultimately, in the lower panel, we find that $\sigma_{\rm lin}$ has a similar response as in the $h$ case, also implying the necessity of doing a joint likelihood analysis including cosmology in results for, e.g., kSZ.

%%%%%%%%%%%%%%%%%%%%%%%%%%%%%%%%%%%%%%%%%%%%%%%%%%%%%%%
\subsubsection{Galaxy linear bias: almost a pure amplitude}

In the absence of RSD, the linear bias $b$ would simply be a multiplicative factor for the reconstructed velocity.
It would therefore have no impact on $r$, and the standard deviation would simply scale as $b^{\rm fid.}/b^w$.
This is the main trend we see on the right panels of Fig. \ref{fig:cosmology}.

In the presence of RSD, an incorrect assumed $b$ does lead to a slight anisotropy in the weighting of Fourier modes in the reconstruction, as shown in App.~\ref{sec:changing_b}.
Indeed, Eq. \ref{eq:factor_for_b} predicts that
$\sigma_{v^{\perp}_{\rm rec}} > \sigma_{v^{\|}_{\rm rec}}$
for $b^w<b^{\rm fid.}$,
as seen in the bottom right panel.

%%%%%%%%%%%%%%%%%%%%%%%%%%%%%%%%%%%%%%%%%%%%%%%%%%%%%%%

\subsection{Photometric-only reconstruction outperforms the hybrid spectro-photometric reconstruction}
\label{sec:hybrid}

As shown in Sec. \ref{sec:perf_metrics}, the kSZ SNR scales as $r \sqrt{N_\text{gal}}$.
Therefore, it is natural to consider photometric samples, denser than the available spectroscopic ones, despite the loss in $r$ from their inaccurate redshifts.
A tempting approach is to use both, spectroscopic and photometric data, in different steps of the velocity reconstruction:
First, to derive the 3D velocity density field from the spectroscopic galaxies, thus without photo-$z$ errors, and second, to then evaluate it at the positions of the much more numerous photometric galaxies.
This approach has been successfully implemented in, e.g. \cite{Mallaby_Kay_2023}.
In this section, we assess whether this hybrid photometric-spectroscopic reconstruction is actually better than simply reconstructing the velocities from the photometric sample directly.

For practical reasons, we choose $\overline{n}_{\rm photo} = 2.14 \cdot 10^{-4}$ Mpc$^{-3}$ (the same as the whole box) and $\overline{n}_{\rm spectra} \sim 4.28 \cdot 10^{-5}$ Mpc$^{-3}$, which is still in the regime that the reconstruction method is unaffected by the decrement of galaxies at fixed volume, as shown in Fig. \ref{fig:density}.
While realistic photometric surveys are generally denser than this, this only reinforces our final conclusion.
In Fig.~\ref{fig:hybrid_r} we show the correlation coefficient of all possible combinations of samples (labeled sample$_1$-sample$_2$) when introducing them into the density and positions stages of the reconstruction method respectively.

Our key finding is that the photo.-photo. case leads to a larger correlation coefficient $r$ than the spect.-photo one.
Thus the photo.-photo. case leads to a higher kSZ SNR.
One may have expected that using the spectroscopic data at the density estimation step should help, since these galaxies have more accurate LOS positions.
This would lead to preserving the small-scale information in the 3D density field, which are otherwise erased in the photo.-photo. case.
However, both cases do the same velocity evaluation step with the photometric galaxies, where photo-$z$ errors place them at the wrong positions.
This leads to misevaluating the small-scale velocity modes in both cases.
So it is better to null these small-scale modes at the density estimation step, which the photo.-photo. case does, rather than keeping these small scale modes like the spect.-photo. case does.
This lowers the variance of the reconstructed velocities, while keeping the covariance with the true velocity unchanged, thus improving the correlation coefficient.
This interpretation matches also what we find on the covariance and standard deviation in App.~\ref{sec:hybrid_extra}, where we explain this result in more detail.

Finally, we have not explored modifications to the reconstruction algorithm to take into account photo-$z$ errors more optimally, or to perform a more optimal joint reconstruction using photometric and spectroscopic samples.
We leave these to future work.

%%%%%%%%%%%%%%%%%%%%%%%%%%%%%%%%%%%%%%%%%%%%%%%%%%%%%%%
%%%%%%%%%%%%%%%%%%%%%%%%%%%%%%%%%%%%%%%%%%%%%%%%%%%%%%%
\section{Conclusions}
\label{sec:conclusions}

In this paper, we have studied the peculiar velocities of galaxies and their host halos, inferred from the Zel'dovich approximation in an \textsc{AbacusSummit} simulation box.
We thoroughly studied the velocity reconstruction performance through two metrics, the correlation coefficient and the standard deviation, that are key inputs for the signal-to-noise and the amplitude of stacked kSZ and moving lens measurements. 
We study idealized cubic periodic 3D boxes, with realistic galaxy selection for a DESI or Rubin-like survey.
Our study finds the following key results:
\begin{itemize}
    \item For kSZ and the moving lens effect, the halo velocities, rather than the galaxy velocities, are what we really seek to reconstruct. Not only are halo velocities not affected by the virial motion of satellite galaxies, but these virial motions are not probed by standard kSZ or moving lens measured (other than, e.g., rotating kSZ).
    This realization shows that the correlation coefficient between reconstructed and true halo velocities is actually slightly larger ($r=0.73$) than the previously considered one between reconstructed and true galaxy velocities ($r=0.55$) along the LOS.
    This difference will matter for the upcoming high-SNR measurements with ACT/SO and DESI/Rubin.    
    
    \item While photometric redshifts degrade the velocity reconstruction by order unity, the increase in sample size available with the next generation of photometric surveys more than compensates for this.
    \item The velocity reconstruction performance saturates for galaxy number densities $\overline{n}_\text{gal} \gtrsim 5 \cdot 10^{-5}$, corresponding to current spectroscopic surveys.
    Beyond this value, an increment of galaxies in a fixed volume cannot be utilized by the reconstruction method we adopt: namely, the first-order Zel'dovich approximation.
    
    \item The satellite fraction is not a crucial parameter for the performance of reconstructing halo velocities.
    We find the correlation coefficient to be mostly insensitive to varying satellite fraction at fixed number density for a single HOD realization.
    This would not be true if we compared the reconstructed velocities with the galaxy (instead of halo) velocities: there, a higher satellite fraction reduces the correlation coefficient $r$.

    \item We did not vary the amplitude of the random virial motions for central ($\alpha_c$) and satellite ($\alpha_s$) galaxies. However, for the purpose of velocity reconstruction, increasing them should be identical to increasing the satellite fraction, since both lead to a galaxy sample with higher random motions.
    As discussed above, this has a minimal effect on the reconstruction of halo velocities.

    \item There is an optimal smoothing scale of the galaxy number density and it is different for the components across and along the LOS ($r^{\perp}_s = 5.9$ Mpc and $r^{\|}_s = 14.8$ Mpc).
    When the smoothing scale is smaller than the optimal value, shot noise dominates, while on larger smoothing scales, useful large-scale density modes are erased and therefore the reconstruction is worsened.
    
    \item Assuming incorrect cosmological parameters in the velocity reconstruction affects it at the $\sim$ percent level.
    This may become relevant for upcoming kSZ and moving lens measurements.
    We sketch how a consistent joint analysis can be performed, in order to propagate these uncertainties into gas profile uncertainties, given a kSZ measurement.
    The Hubble parameter ($h$) acts as an isotropic re-scaling of the simulated box, while the matter density ($\Omega_m^0$) induces a more complex effect, which combines an anisotropic re-scaling and impacts the growth rate.
    The linear bias ($b$) acts almost as a multiplicative factor, but the presence of RSD induces an anisotropic behaviour across and along the LOS.
    While some of these effects could be sub-percent, there are still cases in which their wrong inference could impact subsequent results from kSZ measurement (specially $b$), and therefore, we suggest jointly fitting for, e.g. galaxy formation and cosmology.
    
    \item We find that the simplest hybrid spectroscopic-photometric velocity reconstruction leads to a worse result than doing an analysis only with photometric data in terms of the correlation coefficient, and therefore, the kSZ signal-to-noise. 
    Additionally, we identify and characterize the steps at which the reconstruction is deteriorated when using either spectroscopic and/or photometric data.
\end{itemize}

Our work focuses on idealized cubic periodic boxes at a fixed redshift (with a realistic galaxy sample) in order to isolate most easily the various effects above.
However, for realistic values of the correlation coefficient and standard deviation of the reconstructed velocities, a realistic analysis on a light cone is most desirable.
This, along with a realistic survey mask and survey selection function, are considered in the companion paper \cite{Hadzhiyska2023}, where we derive our most realistic estimates of the velocity reconstruction performance for DESI-like and Rubin-like galaxies.

%%%%%%%%%%%%%%%%%%%%%%%%%%%%%%%%%%%%%%%%%%%%%%%%%%%%%%%

\section*{Acknowledgments}

We thank Tom Abel, Alexander Roman, Kendrick Smith, Sihan Yuan, Risa Wechsler, David Valcin and Martin White for useful discussions. 
This work received support from the U.S. Department of Energy under contract number DE-AC02-76SF00515 to SLAC National Accelerator Laboratory.
This research used resources of the National Energy Research Scientific Computing Center (NERSC), a U.S. Department of Energy Office of Science User Facility located at Lawrence Berkeley National Laboratory.
B.H. is generously supported by the Miller Institute at University of California, Berkeley.
S.F. is supported by Lawrence Berkeley National Laboratory and the Director, Office of Science, Office of High Energy Physics of the U.S. Department of Energy under Contract No.\ DE-AC02-05CH11231.

%%%%%%%%%%%%%%%%%%%%%%%%%%%%%%%%%%%%%%%%%%%%%%%%%%%%%%%
\bibliographystyle{prsty.bst}
\bibliography{refs}

\begin{thebibliography}{10}

\bibitem{Hadzhiyska2023}
B. Hadzhiyska, S. Ferraro, B. Ried~Guachalla, and E. Schaan, Phys. Rev. D {\bf
  109},  103534  (2024).

\bibitem{Poincare1899}
H. Poincar{\'e} and R. Magini, Il Nuovo Cimento (1895-1900) {\bf 10},  128
  (1899).

\bibitem{Arnold1976}
A.~V. Igorevich, E. Djilali, and V.~I. Arnold, {\em Les méthodes
  mathématiques de la mécanique classique} (Éditions Mir, Moscou, C 1976).

\bibitem{Perryman1997}
M.~A.~C. {Perryman} {\it et~al.}, \aap {\bf 323},  L49  (1997).

\bibitem{Bailer_Jones_2015}
C.~A.~L. Bailer-Jones, Publications of the Astronomical Society of the Pacific
  {\bf 127},  994  (2015).

\bibitem{Leavitt1912}
H.~S. {Leavitt} and E.~C. {Pickering}, Harvard College Observatory Circular
  {\bf 173},  1  (1912).

\bibitem{Lemaitre1927}
G. {Lema{\^\i}tre}, Annales de la Soci\&eacute;t\&eacute; Scientifique de
  Bruxelles {\bf 47},  49  (1927).

\bibitem{Hubble1929}
E. {Hubble}, Proceedings of the National Academy of Science {\bf 15},  168
  (1929).

\bibitem{Sandage1958}
A. {Sandage}, \apj {\bf 127},  513  (1958).

\bibitem{Riess_1995}
A.~G. Riess, W.~H. Press, and R.~P. Kirshner, The Astrophysical Journal {\bf
  438},  L17  (1995).

\bibitem{Kaiser1987}
N. {Kaiser}, \mnras {\bf 227},  1  (1987).

\bibitem{Strauss1995}
M.~A. {Strauss} and J.~A. {Willick}, \physrep {\bf 261},  271  (1995).

\bibitem{Howlett_2022}
C. Howlett {\it et~al.}, Monthly Notices of the Royal Astronomical Society {\bf
  515},  953  (2022).

\bibitem{Percival_2009}
W.~J. Percival and M. White, Monthly Notices of the Royal Astronomical Society
  {\bf 393},  297  (2009).

\bibitem{smith2018ksz}
K.~M. Smith {\it et~al.}, KSZ tomography and the bispectrum, 2018.

\bibitem{Munchmeyer_2019}
M. Münchmeyer {\it et~al.}, Constraining local non-Gaussianities with kinetic
  Sunyaev-Zel'dovich tomography, 2019.

\bibitem{Peebles_2022}
P. Peebles, Annals of Physics {\bf 447},  169159  (2022).

\bibitem{Grishchuk1978}
L.~P. {Grishchuk} and I.~B. {Zeldovich}, \sovast {\bf 22},  125  (1978).

\bibitem{Tully1977}
R.~B. {Tully} and J.~R. {Fisher}, \aap {\bf 54},  661  (1977).

\bibitem{1987Djorgovski}
S. {Djorgovski} and M. {Davis}, \apj {\bf 313},  59  (1987).

\bibitem{Branch1993}
D. {Branch} and D.~L. {Miller}, \apjl {\bf 405},  L5  (1993).

\bibitem{Sunyaev1980}
R.~A. {Sunyaev} and Y.~B. {Zeldovich}, \mnras {\bf 190},  413  (1980).

\bibitem{Fukugita_2004}
M. Fukugita and P.~J.~E. Peebles, The Astrophysical Journal {\bf 616},  643
  (2004).

\bibitem{Bullock2017}
J.~S. {Bullock} and M. {Boylan-Kolchin}, \araa {\bf 55},  343  (2017).

\bibitem{Battaglia_2017}
N. Battaglia, S. Ferraro, E. Schaan, and D.~N. Spergel, Journal of Cosmology
  and Astroparticle Physics {\bf 2017},  040  (2017).

\bibitem{Walker1991}
T.~P. {Walker} {\it et~al.}, \apj {\bf 376},  51  (1991).

\bibitem{White1993}
S.~D.~M. {White}, J.~F. {Navarro}, A.~E. {Evrard}, and C.~S. {Frenk}, \nat {\bf
  366},  429  (1993).

\bibitem{Planck2020}
{Planck Collaboration} {\it et~al.}, \aap {\bf 641},  A6  (2020).

\bibitem{Cen_2006}
R. Cen and J.~P. Ostriker, The Astrophysical Journal {\bf 650},  560  (2006).

\bibitem{1998Bryan}
G.~L. {Bryan} and M.~L. {Norman}, \apj {\bf 495},  80  (1998).

\bibitem{2009Sun}
M. {Sun} {\it et~al.}, \apj {\bf 693},  1142  (2009).

\bibitem{2020Macquart}
J.~P. {Macquart} {\it et~al.}, \nat {\bf 581},  391  (2020).

\bibitem{2018Nicastro}
F. {Nicastro} {\it et~al.}, \nat {\bf 558},  406  (2018).

\bibitem{lsstsciencecollaboration2009lsst}
L.~S. Collaboration {\it et~al.}, LSST Science Book, Version 2.0, 2009.

\bibitem{Amendola_2018}
L. Amendola {\it et~al.}, Cosmology and fundamental physics with the Euclid
  satellite, 2018.

\bibitem{Spergel_2015}
D. {Spergel} {\it et~al.}, {Wide-Field InfrarRed Survey Telescope-Astrophysics
  Focused Telescope Assets WFIRST-AFTA 2015 Report}, 2015.

\bibitem{Birkinshaw1983}
M. {Birkinshaw} and S.~F. {Gull}, \nat {\bf 302},  315  (1983).

\bibitem{Hotinli2019}
S.~C. {Hotinli} {\it et~al.}, \prl {\bf 123},  061301  (2019).

\bibitem{Hotinli2021}
S.~C. {Hotinli}, K.~M. {Smith}, M.~S. {Madhavacheril}, and M. {Kamionkowski},
  \prd {\bf 104},  083529  (2021).

\bibitem{Ade2019}
P. {Ade} {\it et~al.}, \jcap {\bf 2019},  056  (2019).

\bibitem{Abazajian2016}
K.~N. {Abazajian} {\it et~al.}, {CMB-S4 Science Book, First Edition}, 2016.

\bibitem{Aghamousa2016}
{DESI Collaboration} {\it et~al.}, {The DESI Experiment Part I:
  Science,Targeting, and Survey Design}, 2016.

\bibitem{Zeldovich1970}
Y.~B. {Zel'dovich}, \aap {\bf 5},  84  (1970).

\bibitem{Eisenstein_2007}
D.~J. {Eisenstein}, H.-J. {Seo}, E. {Sirko}, and D.~N. {Spergel}, \apj {\bf
  664},  675  (2007).

\bibitem{Padmanabhan_2012}
N. Padmanabhan {\it et~al.}, Monthly Notices of the Royal Astronomical Society
  {\bf 427},  2132  (2012).

\bibitem{Fowler2007}
J.~W. {Fowler} {\it et~al.}, \ao {\bf 46},  3444  (2007).

\bibitem{Swetz2011}
D.~S. {Swetz} {\it et~al.}, \apjs {\bf 194},  41  (2011).

\bibitem{Thornton2016}
R.~J. {Thornton} {\it et~al.}, \apjs {\bf 227},  21  (2016).

\bibitem{Maksimova_2021}
N.~A. Maksimova {\it et~al.}, Monthly Notices of the Royal Astronomical Society
  {\bf 508},  4017  (2021).

\bibitem{Garrison2019}
L.~H. {Garrison}, D.~J. {Eisenstein}, and P.~A. {Pinto}, \mnras {\bf 485},
  3370  (2019).

\bibitem{Garrison2021}
L.~H. {Garrison} {\it et~al.}, \mnras {\bf 508},  575  (2021).

\bibitem{DESIReport}
{DESI Collaboration} {\it et~al.}, {The DESI Experiment Part I:
  Science,Targeting, and Survey Design}, 2016.

\bibitem{Eisenstein_2011}
D.~J. Eisenstein {\it et~al.}, The Astronomical Journal {\bf 142},  72  (2011).

\bibitem{Dawson_2012}
K.~S. Dawson {\it et~al.}, The Astronomical Journal {\bf 145},  10  (2012).

\bibitem{2023Zhou}
R. {Zhou} {\it et~al.}, \aj {\bf 165},  58  (2023).

\bibitem{Yuan2023}
S. {Yuan} {\it et~al.}, {The DESI One-Percent Survey: Exploring the Halo
  Occupation Distribution of Luminous Red Galaxies and Quasi-Stellar Objects
  with AbacusSummit}, 2023.

\bibitem{2010Tinker}
J.~L. {Tinker} {\it et~al.}, \apj {\bf 724},  878  (2010).

\bibitem{Yuan_2021}
S. Yuan {\it et~al.}, Monthly Notices of the Royal Astronomical Society {\bf
  510},  3301–3320  (2021).

\bibitem{Hadzhiyska2022a}
B. {Hadzhiyska} {\it et~al.}, \mnras {\bf 525},  4367  (2023).

\bibitem{Yuan_2023_Placeholder}
S. {Yuan} {\it et~al.}, {(in prep.)}, 2023.

\bibitem{2007Zheng}
Z. {Zheng}, A.~L. {Coil}, and I. {Zehavi}, \apj {\bf 667},  760  (2007).

\bibitem{2015Guo}
H. {Guo} {\it et~al.}, \mnras {\bf 446},  578  (2015).

\bibitem{2017Ye}
J.-N. {Ye}, H. {Guo}, Z. {Zheng}, and I. {Zehavi}, \apj {\bf 841},  45  (2017).

\bibitem{hadzhiyska2023synthetic}
B. Hadzhiyska {\it et~al.}, Synthetic light cone catalogues of modern redshift
  and weak lensing surveys with AbacusSummit, 2023.

\bibitem{yuan2023desi}
S. Yuan {\it et~al.}, The DESI One-Percent Survey: Exploring the Halo
  Occupation Distribution of Luminous Red Galaxies and Quasi-Stellar Objects
  with AbacusSummit, 2023.

\bibitem{Diego_Blas_2011}
D. Blas, J. Lesgourgues, and T. Tram, Journal of Cosmology and Astroparticle
  Physics {\bf 2011},  034–034  (2011).

\bibitem{Takahashi_2012}
R. Takahashi {\it et~al.}, The Astrophysical Journal {\bf 761},  152  (2012).

\bibitem{Desjacques2018}
V. {Desjacques}, D. {Jeong}, and F. {Schmidt}, \physrep {\bf 733},  1  (2018).

\bibitem{White2015}
M. {White}, \mnras {\bf 450},  3822  (2015).

\bibitem{chan2023reconstructing}
K.~C. Chan, G. Lu, and X. Wang, Reconstructing the Baryonic Acoustic
  Oscillations in the presence of photo-$z$ uncertainties, 2023.

\bibitem{Schaan2021}
E. Schaan {\it et~al.}, Phys. Rev. D {\bf 103},  063513  (2021).

\bibitem{Sheth_2001}
R.~K. Sheth and A. Diaferio, Monthly Notices of the Royal Astronomical Society
  {\bf 322},  901–917  (2001).

\bibitem{Takada_2014}
M. Takada {\it et~al.}, Publications of the Astronomical Society of Japan {\bf
  66},  R1  (2014).

\bibitem{LSST_SB_2009}
{LSST Science Collaboration} {\it et~al.}, {LSST Science Book, Version 2.0},
  2009.

\bibitem{Eisenstein2005}
D.~J. {Eisenstein} {\it et~al.}, \apj {\bf 633},  560  (2005).

\bibitem{Reid2016}
B. {Reid} {\it et~al.}, \mnras {\bf 455},  1553  (2016).

\bibitem{rocher2023desi}
A. Rocher {\it et~al.}, The DESI One-Percent survey: exploring the Halo
  Occupation Distribution of Emission Line Galaxies with AbacusSummit
  simulations, 2023.

\bibitem{yuan2023unraveling}
S. Yuan {\it et~al.}, Unraveling emission line galaxy conformity at z~1 with
  DESI early data, 2023.

\bibitem{Riess_2022}
A.~G. Riess {\it et~al.}, The Astrophysical Journal Letters {\bf 934},  L7
  (2022).

\bibitem{Mallaby_Kay_2023}
M. Mallaby-Kay {\it et~al.}, Kinematic Sunyaev-Zel’dovich effect with ACT,
  DES, and BOSS: A novel hybrid estimator, 2023.

\end{thebibliography}

%%%%%%%%%%%%%%%%%%%%%%%%%%%%%%%%%%%%%%%%%%%%%%%%%%%%%%%
%%%%%%%%%%%%%%%%%%%%%%%%%%%%%%%%%%%%%%%%%%%%%%%%%%%%%%%

\appendix

%%%%%%%%%%%%%%%%%%%%%%%%%%%%%%%%%%%%%%%%%%%%%%%%%%%%%%%

\section{Velocity dispersion across the line-of-sight}
\label{app:across_LOS}

\begin{figure}[H]
     \centering
     \begin{subfigure}[b]{0.48\textwidth}
         \centering
         \includegraphics[width=0.99\textwidth]{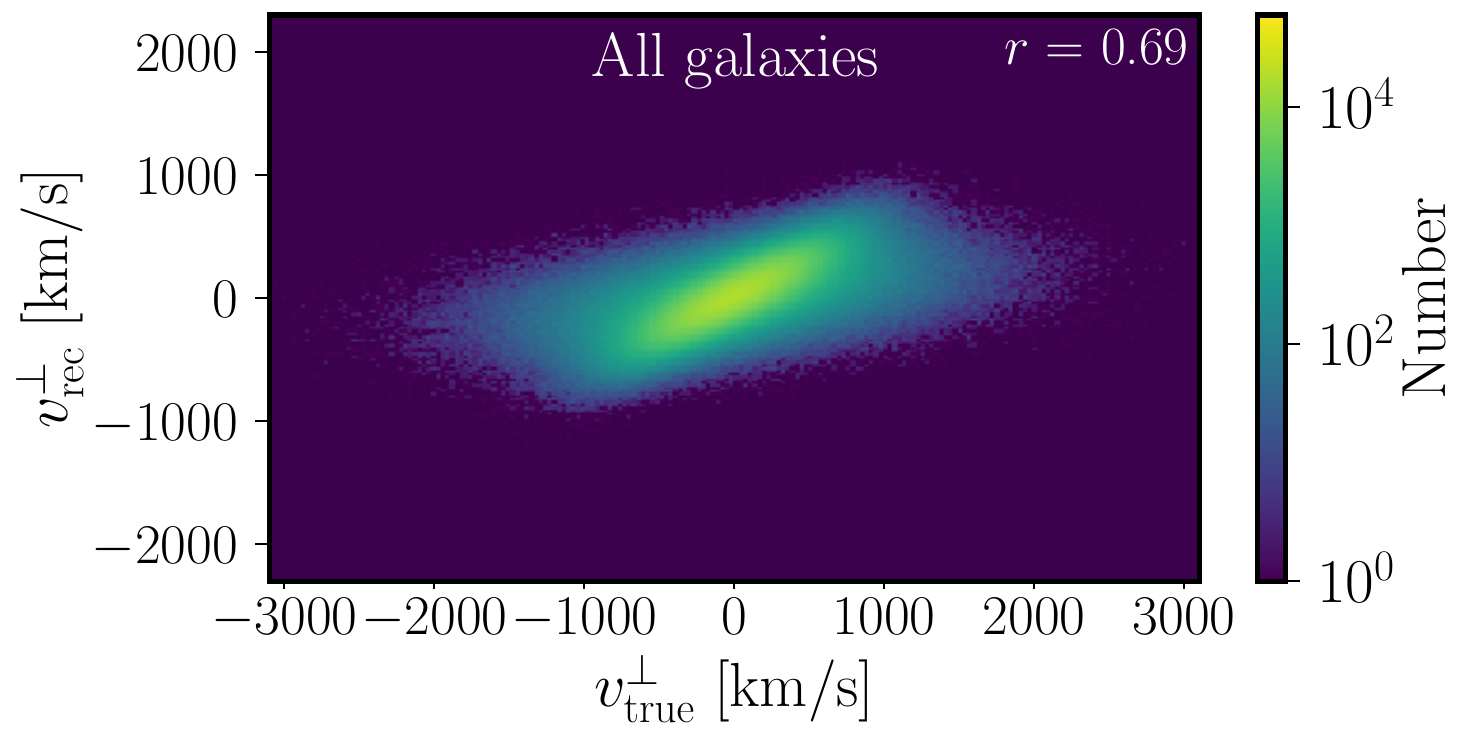}
     \end{subfigure}
     \hfill
     \begin{subfigure}[b]{0.48\textwidth}
         \centering
         \includegraphics[width=0.99\textwidth]{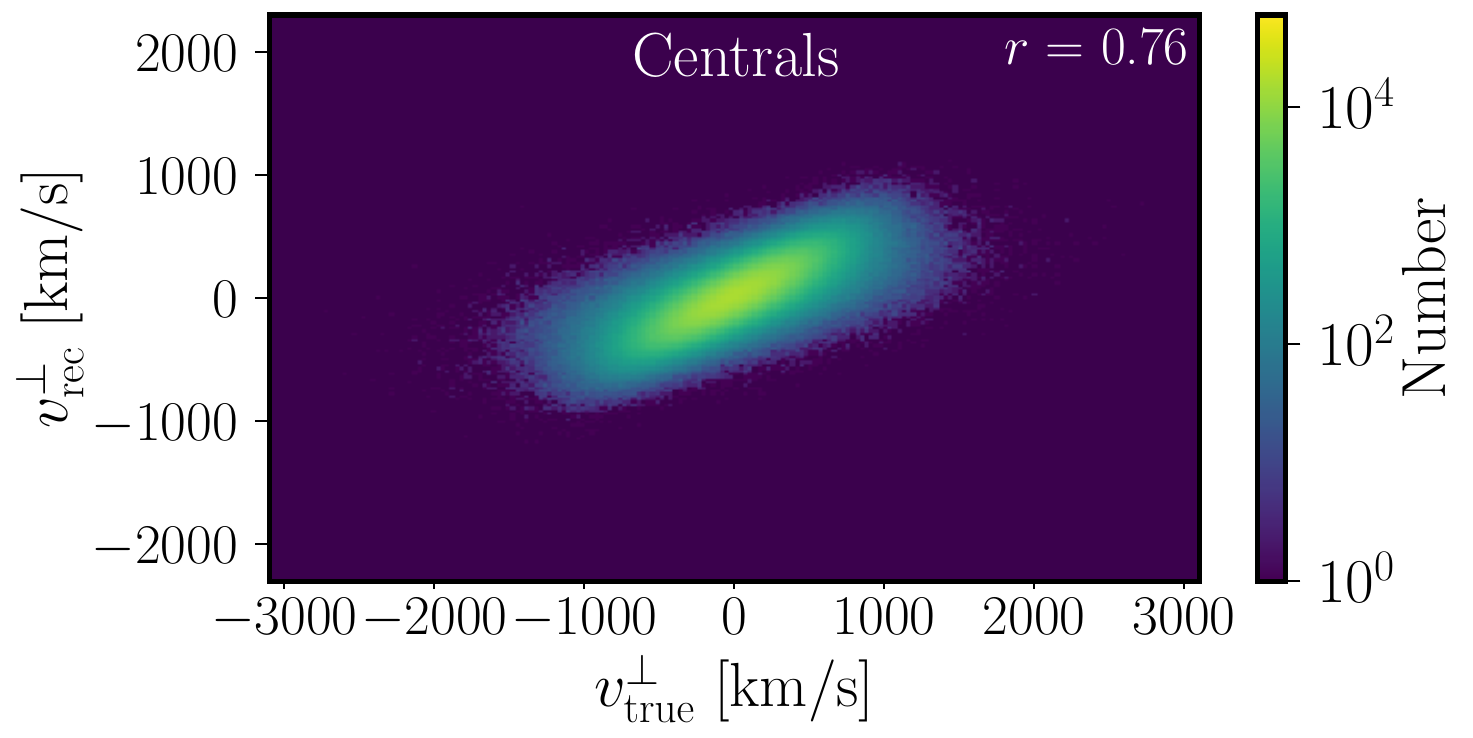}
     \end{subfigure}
     \hfill
     \begin{subfigure}[b]{0.48\textwidth}
         \centering
         \includegraphics[width=0.99\textwidth]{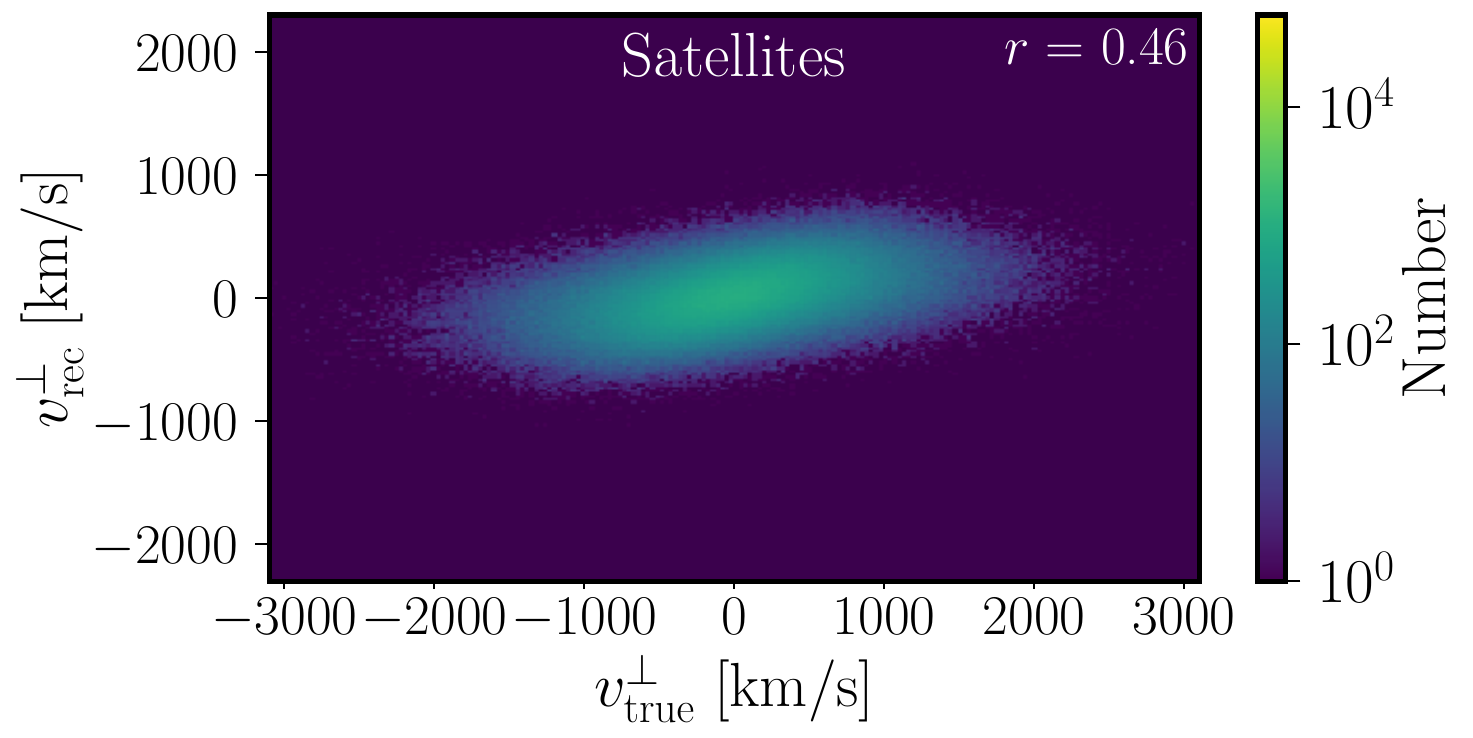}
     \end{subfigure}
        \caption{
        Similar to Fig. \ref{fig:vrec_vcent_vsat_histogram_los}, but rather than along it, across the line-of-sight projection.
        Reconstruction improves in all ranges due to the absence of redshift space distortions.
        For all ranges, $v^{\perp}_{\rm rec}$ is positively correlated with $v^{\perp}_{\rm true}$, but larger velocities are underestimated.
        }
    \label{fig:vrec_vcent_vsat_histogram_across}
\end{figure}

\begin{figure}[H]
    \centering
    \includegraphics[width=0.48\textwidth]{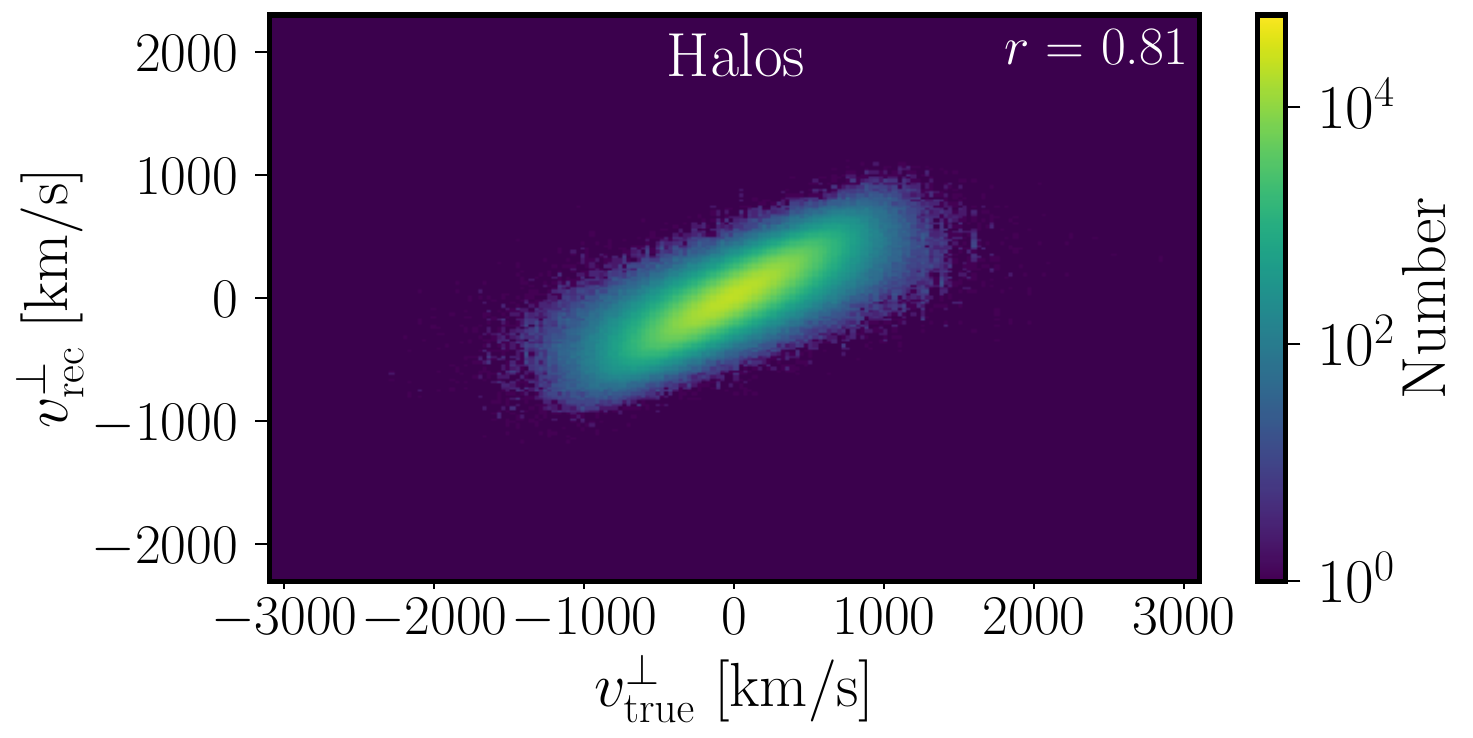}
    \caption{
        The 2D histograms of the halo velocity reconstruction across the line-of-sight. }
    \label{fig:vrec_vhalos_2D_histogram_across}
\end{figure}

As mentioned in Sec. \ref{sec:vel_rec}, there are two components of the velocity reconstructed field: across and along the LOS.
Fig. \ref{fig:vrec_vhalos_2D_histogram_across} shows the 2D log-histogram distribution of $v^{\perp}_{\rm true}$ and $v^{\perp}_{\rm gal}$, similarly to Fig. \ref{fig:vrec_vhalos_2D_histogram_los}, but for the across the LOS component. \\

The complete sample of galaxies across the LOS have a better performance than the case along the LOS (improves from 0.55 to 0.69).
When splitting between centrals and satellites, we again find that the performance on centrals is better due to the lower contribution of thermal velocities.
The overall correlation is higher because RSD is not as relevant when studying the across the LOS contribution. \\

In Fig. \ref{fig:vrec_vhalos_2D_histogram_across} we show the corresponding 2D log-histogram for the across the LOS velocity of the halos, similarly to Fig. \ref{fig:vrec_vhalos_2D_histogram_los}.
The correlation coefficient reaches the best performance of this work: $r^{\perp}_{\rm halos} = 0.81$.
This improvement is promising for detecting the moving lens effect in future cosmological surveys.

%%%%%%%%%%%%%%%%%%%%%%%%%%%%%%%%%%%%%%%%%%%%%%%%%%%%%%%

\section{Algebraic derivation of the impact of incorrect cosmological parameters}
\label{sec:algebra_wrong_params}

Previously, in Sec. \ref{sec:wrong_cosmo}, we present the effect on the reconstruction of velocities when assuming cosmological parameters different from the fiducial ones. 
In this section, we derive the underlying algebra in order to get intuition and a correct interpretation. \\

This appendix is structured as follows:
In Sec. \ref{sec:spatial_coord} we present the general framework and notation, in addition to the spatial conversion from positions in a cubic box to positions in the sky.
In the following sections (Sec. \ref{sec:changing_h}, Sec. \ref{sec:changing_omega}, and Sec. \ref{sec:changing_b}) we derive the algebra for $h$, $\Omega_m$ and $b$ respectively.

%%%%%%%%%%%%%%%%%%%%%%%%%%%%%%%%%%%%%%%%%%%%%%%%%%%%%%%

\subsection{Re-scaling of spatial coordinates}
\label{sec:spatial_coord}

On the cubic box we use 3D spatial coordinates to describe the positions of galaxies.
Translating this into a real-scenario of a galaxy catalog, we instead have two angles:
right ascension (RA) and declination (DEC),
and a redshift $z$.
Given RA, DEC and $z$ for a galaxy, one could infer the 3D Cartesian position as follows:
\beq
\left\{
\bal
    &{\rm RA} \approx \frac{x}{\chi} \\
    &{\rm DEC} \approx \frac{y}{\chi} \\
    &\Delta z \approx \frac{\Delta \chi H}{c} 
\eal
\right.
,
\eeq
where $H$ is the Hubble parameter.
In a flat matter-dominated Universe, this is:
\beq
H(z) = h \cdot 100 \cdot \sqrt{\Omega_m(1+z)^3 + (1-\Omega_m)} ,
\eeq
and $\chi$ is the co-moving distance:
\beq
\chi(z) = \int_0^z \dfrac{c \cdot dz'}{H(z')} ,
\eeq
with a fixed redshift $H(z=0.5)$ and $\chi(z=0.5)$. \\

A change on the cosmological parameters impacts the 3D Cartesian positions of galaxies when fixing the celestial coordinates and redshifts as follows:
\beq
\left\{
\bal
    &x_{\perp, w} = \frac{\chi_w}{\chi_t} x_{\perp, t} \\
    &x_{\parallel, w} = \frac{H_t}{H_w} x_{\parallel, t}
\eal
\right. 
,
\eeq
where $x_{\perp}$ and $x_{\|}$ correspond to the coordinates across and along the LOS, while the subscript $w$ and $t$ correspond to \textit{wrong} and \textit{true} (i.e. fiducial) parameters.
In Fourier space, this translates to:
\beq
\left\{
\bal
    &k_{\perp, w} = \frac{\chi_t}{\chi_w} k_{\perp, t} \\
    &k_{\parallel, w} = \frac{H_w}{H_t} k_{\parallel, t} \\
    & k_w = \sqrt{\left( \frac{\chi_t}{\chi_w} \right)^2 k_{\perp, t}^2 + \left( \frac{H_w}{H_t} \right)^2 k_{\parallel, t}^2} \\
    &\mu_{t, w} = \frac{k_{\parallel, t, w}}{k_{t, w}}\\
\eal
\right.
\eeq
\\

Thus, for any scalar function $g_t(x_t) = g_w(x_w)$ in configuration space, we have:
\beq
    g_w(\mathbf{k}_w) =
    \left( \frac{\chi_w}{\chi_t} \right)^2
    \left( \frac{H_t}{H_w} \right)
    g_t \left( \mathbf{k}_t\right).
\eeq
Similarly, for any scalar function $g_t(k_t) = g_w(k_w)$ in configuration space, we get:
\beq
    g_w(\mathbf{x}_w) =
    \left( \frac{\chi_t}{\chi_w} \right)^2
    \left( \frac{H_w}{H_t} \right)
    g_t \left( \mathbf{x}_t\right).
\eeq

%%%%%%%%%%%%%%%%%%%%%%%%%%%%%%%%%%%%%%%%%%%%%%%%%%%%%%%

\subsection{Changing \texorpdfstring{$h$}{Lg}}
\label{sec:changing_h}

The only effect of assuming an incorrect value of $h$ is a re-scaling of the Cartesian coordinates as above, with 
$\chi_w / \chi_t
=
H_t / H_w
=
h_t / h_w$.
In the absence of a fixed smoothing scale, this re-scaling turns out to cancel with the additional amplitude factors in the velocity reconstruction.
However, given that our smoothing scale is fixed in Cartesian space to 14.8 Mpc, it does not get re-scaled by $h_t / h_w$, thus breaking this cancellation.
In other words, in terms of impact on the reconstructed velocities, changing $h$ is equivalent to re-scaling only the smoothing scale. \\

Indeed, in Fourier space,
\beq
\bal
\mathbf{v}_w(\mathbf{k}_w, r_s) 
&=\mathbf{v}_w(\mathbf{k}_w) \;
W(k_w, r_s) \\
&= a H_w f \; 
\frac{i \mathbf{k}_w}{k_w^2}  \;
\frac{\delta^g_w(\mathbf{k}_w)}{\left( b + f \mu_w^2 \right)} \;
W(k_w, r_s)\\
&= \left( \frac{h_t}{h_w} \right)^3 \;
a H_t f \;
\frac{i \mathbf{k}_t}{k_t^2} 
\frac{\delta^g_t(\mathbf{k}_t)}{{\left( b + f \mu_t^2 \right)}}  \;
e^{-\frac{1}{2} \left( \frac{h_w}{h_t} k_t \right)^2 r_s^2 }\\
&= \left( \frac{h_t}{h_w} \right)^3  \;
\mathbf{v}_t(\mathbf{k}_t) \;
e^{-\frac{1}{2} k_t^2 \left( \frac{h_w}{h_t} r_s\right)^2 }\\
&= \left( \frac{h_t}{h_w} \right)^3  \;
\mathbf{v}_t(\mathbf{k}_t)
W \left( k_t, \frac{h_w}{h_t} r_s \right) \\
&=
\left( \frac{h_t}{h_w} \right)^3  \;
\mathbf{v}_t \left( \mathbf{k}_t, \frac{h_w}{h_t} r_s \right)
\eal
\eeq
\\
As a result, in configuration space:
\beq
\mathbf{v}_w(\mathbf{x}_w, r_s) 
=
\mathbf{v}_t \left( \mathbf{x}_t, \frac{h_w}{h_t} r_s \right),
\eeq
such that the effect on the reconstructed galaxy velocities simply amounts to re-scaling the smoothing scale as
$r_s^{\rm eff} \equiv \dfrac{h_t}{h_w} r_s$. \\

We confirm this numerically in the left panels of Fig.~\ref{fig:cosmology}, by comparing two approaches.
The blue dots show the effect of performing the reconstruction with $h_w$, while the dashed line is obtained by using $h_t$, but only changing the smoothing scale. \\

%%%%%%%%%%%%%%%%%%%%%%%%%%%%%%%%%%%%%%%%%%%%%%%%%%%%%%%

\subsection{Changing \texorpdfstring{$\Omega_m$}{Lg}}
\label{sec:changing_omega}

Assuming a wrong value on $\Omega_m$ has more effects on the reconstruction method:
First, it re-scales the simulation box differently for the coordinates along and across the LOS, introducing an anisotropy.
Second, because of the anisotropy, the RSD effects introduce an additional distortion, and therefore, the modes across and along the LOS are weighted differently.
Third, in the absence of a change on the smoothing, it will re-scale and also differently along and across the LOS.
Finally, it impacts the growth rate $f$ as:
\begin{equation}
    f \approx \Omega_m(z)^{0.55},
\end{equation}
with 
\begin{equation}
    \Omega_m(z) = \frac{\Omega_m(1+z)^3}{\Omega_m(1+z)^3 + (1-\Omega_m)}.
\end{equation}
The velocities could be described separately across and along the LOS as:
\begin{equation}
    \begin{pmatrix}
    v^{\perp}_w(\mathbf{k}_w)\\
    v^{\|}_w(\mathbf{k}_w)
    \end{pmatrix} 
    = 
    a H_w f_w \frac{1}{k_w^2} \frac{\delta^g_w(\mathbf{k}_w)}{(b + f_w \mu_w^2)} 
    \begin{pmatrix}
    k^{\perp}_w\\
    k^{\|}_w
    \end{pmatrix} 
    W(k_w, r_s) ,
\label{eq:vel_omega_wrong}
\end{equation}
re-organizing terms, and relating some factors with the fiducial parameters, we can rewrite Eq. \ref{eq:vel_omega_wrong} as:
\begin{equation}
    \begin{pmatrix}
    v^{\perp}_w(\mathbf{k}_w)\\
    v^{\|}_w(\mathbf{k}_w)
    \end{pmatrix} 
    =
    F_{\Omega_m}(\mu_t)
    \begin{pmatrix}
    A^{\perp} v^{\perp}_t(\mathbf{k}_t)\\
    A^{\|}
    v^{\|}_t(\mathbf{k}_t)
    \end{pmatrix}  
    W(k_t, r_{s, \rm eff}^{\perp}, r_{s, \rm eff}^{\|})
\end{equation}
with $A^{\perp}$ and $A^{\|}$ constants defined as:
\beq
\bal
    A^{\perp} & = \frac{f_w}{f_t} \left( \frac{\chi_w}{\chi_t} \right)^2 \frac{H_w}{H_t} \\
    A_{\|}    & = \frac{f_w}{f_t} \left( \frac{\chi_w}{\chi_t} \right) ,
\eal
\label{eq:constants_Omega_m}
\eeq
$F_{\Omega_m}(\mu)$ a function defined as:
\begin{equation}
    F_{\Omega_m}(\mu) = \frac{b(1-\mu^2) + (f_t + b) \mu^2}{b \frac{\chi_t}{\chi_w} (1-\mu^2) + (f_w + b) \frac{H_w}{H_t} \mu^2} ,
\end{equation}
and $W(k_t, r_{s, \rm eff}^{\perp}, r_{s, \rm eff}^{\|})$ the anisotropic smoothing kernel with $r_{s, \rm eff}^{\perp} = \frac{H_t}{H_w} r_s$ and $r_{s, \rm eff}^{\|} = \frac{\chi_w}{\chi_t} r_s$. \\
We can analyze the impact of each component independently:
\begin{itemize}
    \item The constant parameters $A^{\perp}$ and $A^{\|}$ do impact differently across and along the LOS magnitude of the reconstructed velocities, and therefore, $\sigma_{\rm rec}$.
    However, these constants do not impact the correlation coefficients $r(v_{\rm rec}^{\perp})$ and $r(v_{\rm rec}^{\|})$. 
    \item The function and smoothing kernel combined $F_{\Omega_m}(\mu_t) W(k_t, r_{s, \rm eff}^{\perp}, r_{s, \rm eff}^{\|})$ acts as a new effective smoothing kernel.
    It is different from simply having $W(k_t, s)$ in two ways: \\
    First, $W(k_t, r_{s, \rm eff}^{\perp}, r_{s, \rm eff}^{\|})$ weights $k_w$ differently due to the anisotropic expansion of the box.
    This effect can either increase of decrease $r$, based on Fig. \ref{fig:smoothing} where we vary the smoothing scale. \\
    Second, the combination $F_{\Omega_m}(\mu_t) W(k_t, r_{s, \rm eff}^{\perp}, r_{s, \rm eff}^{\|})$ re-weights the along the LOS contribution $\mu_t$ differently.
    This can only reduce the overall correlation coefficient $r$ in any case.
\end{itemize}

 We confirm that these effects are combined in the central panels from Fig. \ref{fig:cosmology}, and refer to Sec. \ref{sec:wrong_cosmo} for a detailed explanation.

%%%%%%%%%%%%%%%%%%%%%%%%%%%%%%%%%%%%%%%%%%%%%%%%%%%%%%%

\subsection{Changing \texorpdfstring{$b$}{Lg}}
\label{sec:changing_b}

Introducing a value of $b$ different from the fiducial one does not re-scale the simulation box as it is done by the previous parameters. 
We find that the impact on the reconstruction method is directly linked to a re-weighting of the modes across and along the LOS:
\begin{equation}
    \mathbf{v}_w(\mathbf{k}, s) = a H f \frac{i \mathbf{k}}{k^2} \frac{\delta_g(\mathbf{k})}{(b_w + f \mu^2)} W(k, s).
\end{equation}
Similarly, as done in the case of assuming a wrong $\Omega_m$, we can rewrite the velocity contribution in Fourier space as:
\begin{equation}
    \mathbf{v}_w(\mathbf{k}, s) = F_{b}(\mu) \mathbf{v}_t(\mathbf{k}, s) ,
\end{equation}
where 

\begin{equation}
    F_{b}(\mu) = \frac{(b_t + f \mu^2)}{(b_w + f \mu^2)}.
\label{eq:factor_for_b}
\end{equation}
We also find a difference on the performance across and along the LOS due to RSD present when having $\mu$ as a variable.
On each mode, the result can be thought of as an anisotropic constant, therefore, impacting the standard deviation $\sigma_{\rm rec}$ but not the correlation coefficient $r$, as shown in the right panels from Fig. \ref{fig:cosmology}.

%%%%%%%%%%%%%%%%%%%%%%%%%%%%%%%%%%%%%%%%%%%%%%%%%%%%%%%

\section{Understanding the hybrid photometric-spectroscopic results}
\label{sec:hybrid_extra}

\begin{figure*}
    \includegraphics[width=0.86\textwidth]{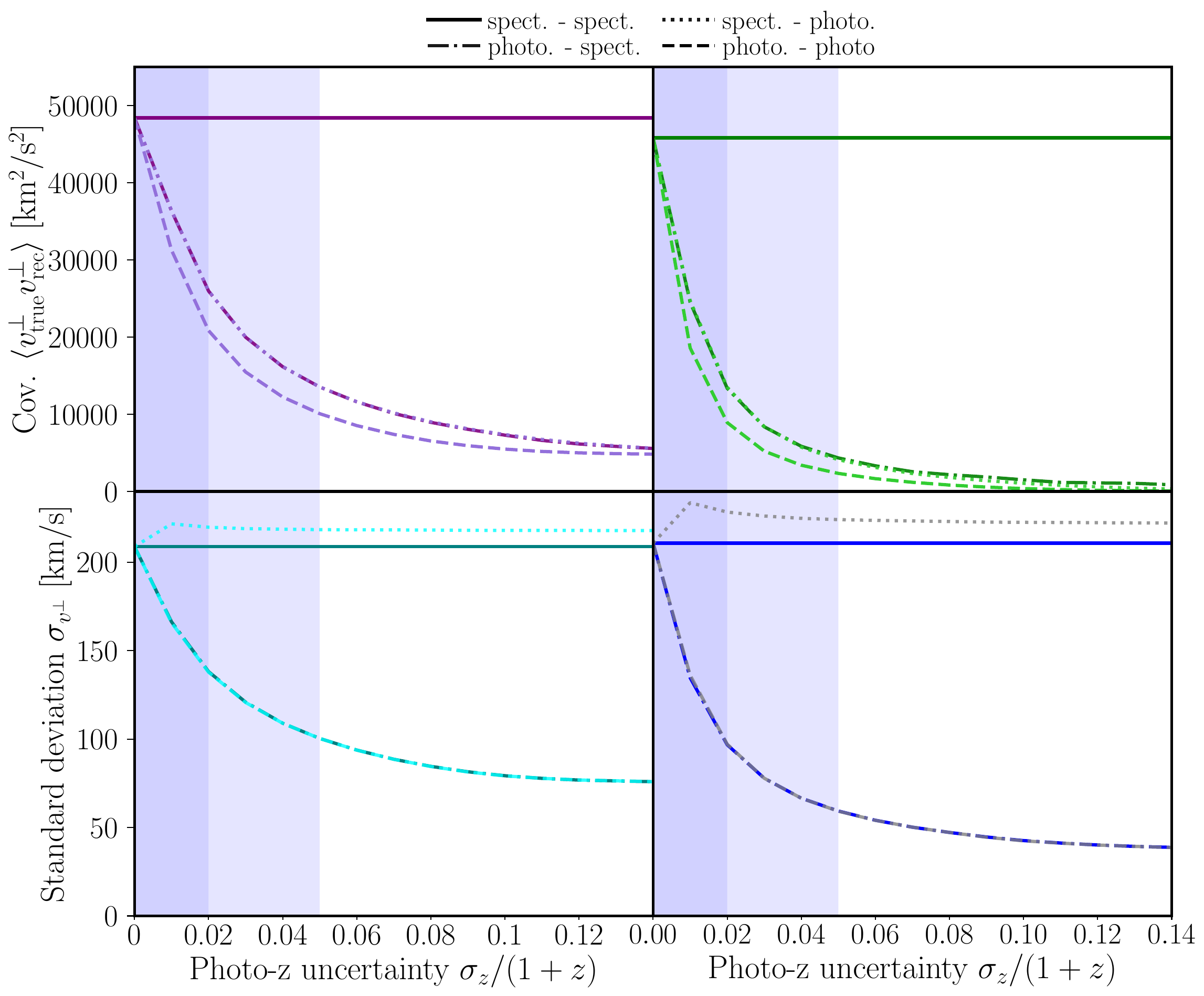}
    \caption{
    Similar to Fig~\ref{fig:hybrid_r}, we plot the covariance and standard deviation for two spectroscopic and photometric samples evaluated in different steps of the reconstruction method transverse to the LOS.
    \textit{Upper panel}: 
    The cases photo.-spect. and spect.-photo. reduce the covariance similarly, but because of different reasons. 
    For photo.-spect., the density field is smoothed due to the photometric noise, removing the small scales velocities.
    For spect.-photo., evaluating the galaxies in wrong positions randomizes the phases of the small scale velocities, and thus reduce the total covariance.
    The photo.-photo. case combined the two effects and therefore, result in the worst case.
    \textit{Bottom panel}: 
    Evaluating the photometric sample at the density stage reduces the standard deviation of the velocities as their amplitude is decreased due to a reduction on their spatial clustering. 
    The standard deviation is the same for photo.-spectra. and photo.-photo. as the positions in which we evaluate the displacements act as a random sample of the density field.
    The case spect.-photo. results in a higher standard deviation as the density field is not smoothed.
    }
    \label{fig:hybrid_cov_std}
\end{figure*}
Photo-z errors have a slightly different effect at the density estimation stage (where one estimates the 3D matter density from the coordinates of the galaxies) and at the velocity estimation stage (where the inferred 3D velocity field is evaluated at the coordinates of the galaxies).
To understand this, it helps to decompose the velocity field as
\begin{equation}
    v = v_{\rm L} + v_{\rm S} ,
\end{equation}
where the subscript L and S stand for large and small scales.
The distinction between large and small is determined by the typical scale of photo-z errors.
This also implies that we can split the covariance 
$C \equiv \langle v^{\rm true} v^{\rm rec} \rangle$
as 
$C = C_{\rm L} + C_{\rm S}$, 
with:
\beq
\bal
    C_{\rm L} &\equiv \langle v^{\rm true} v^{\rm rec}_{\rm L} \rangle  \\ 
    C_{\rm S} &\equiv \langle v^{\rm true} v^{\rm rec}_{\rm S} \rangle
\eal
.
\eeq

Photometric uncertainties at the density estimation stage cause the inferred 3D density field to be convolved with the photo-z error kernel (typically a Gaussian plus tails).
This suppresses the Fourier modes on scales shorter than the typical photo-z uncertainty.
This effectively sets $v^{\text{rec}}_{\rm S} = 0$.

On the other hand, photo-$z$ errors at the velocity estimation stage do not suppress the short-scale Fourier modes, they instead cause them to be evaluated at the wrong positions for the galaxies.
This is equivalent to phase-shifting these Fourier modes:
$v^\text{rec}_{\rm S}$
is replaced by $v^\text{rec shifted}_{\rm S}$.

In both cases, photo-z errors null the correlation between $v^\text{rec}_{\rm S}$ and $v^\text{true}_{\rm S}$,
resulting in $C = C_L$.
Indeed, Fig.~\ref{fig:hybrid_cov_std} shows that both cases (photo.-spect. and spect.-photo.) lead to the same value of $C$, ie the same covariance between true and reconstructed velocities.
And these covariances are almost identical to the photo.-photo. case, as expected from this discussion.

However, while both photo.-spect. and spect.-photo. lead to the same covariance with the true velocities, they lead to different standard deviations of 
$v^{\text{rec}}$.
Indeed, in the photo.-spect. case, the suppression of
$v^\text{rec}_S$
due to the convolution with the photo-$z$ error kernel leads to
$v^{\rm rec} \approx v^{\rm rec}_{\rm L}$.
The same is true in the photo.-photo. case.
On the other hand, as we discussed, the spect.-photo. case does not null $v^\text{rec}_{\rm S}$, it randomizes its phase.
This preserves its variance.
The correspondingly larger variance of $v^\text{true}$ in the spect.-photo. case thus leads to a lower correlation coefficient, as found in Fig.~\ref{fig:hybrid_cov_std}.

All these results combined, lead to the main result find in Sec. \ref{sec:hybrid}, which suggests that using exclusively that photometric sample is better than doing an hybrid spectroscopic-photometric analysis.

\end{document}